\documentclass[AMA,STIX1COL]{WileyNJD-v2}

\articletype{Article Type}%

\received{}
\revised{}
\accepted{}

\usepackage{verbatim} 
\usepackage[listings]{tcolorbox}

\usepackage{mathtools}

\def\b0{{0}}

\newcommand{\ts}{\texttt}

\graphicspath{{images/}} 

\begin{document}

\title{A Flexible Micro-Randomized Trial Design and Sample Size Considerations }

\author[1,2]{Jing Xu*}

\author[1]{Xiaoxi  Yan}

\author[5,6]{Caroline Figueroa}

\author[7,8,9,10]{Joseph Jay Williams}

\author[1,2,3,4]{Bibhas Chakraborty}

\authormark{XU \textsc{et al}}

\address[1]{\orgdiv{Centre for Quantitative Medicine}, \orgname{Duke-NUS Medical School}, \orgaddress{\state{Singapore}, \country{Singapore}}}

\address[2]{\orgdiv{Health Service and Systems Research}, \orgname{Duke-NUS Medical School}, \orgaddress{\state{Singapore}, \country{Singapore}}}

\address[3]{\orgdiv{Department of Statistics and Applied Probability}, \orgname{National University of Singapore}, \orgaddress{\state{Singapore}, \country{Singapore}}}

\address[4]{\orgdiv{Department of Biostatistics and Bioinformatics}, \orgname{Duke University}, \orgaddress{\state{Durham, NC}, \country{USA}}}

\address[5]{\orgdiv{Technology, Policy and Management}, \orgname{Delft University of Technology}, \orgaddress{\state{Delft}, \country{Netherlands}}}

\address[6]{\orgdiv{School of Social Welfare}, \orgname{University of California}, \orgaddress{\state{Berkeley}, \country{USA}}}

\address[7]{\orgdiv{Department of Computer Science}, \orgname{University of Toronto}, \orgaddress{\state{Toronto}, \country{Canada}}}

\address[8]{\orgdiv{Department of Statistical Sciences}, \orgname{University of Toronto}, \orgaddress{\state{Toronto}, \country{Canada}}}

\address[9]{\orgdiv{Department of Psychology}, \orgname{University of Toronto}, \orgaddress{\state{Toronto}, \country{Canada}}}

\address[10]{\orgdiv{Vector Institute for Artificial Intelligence Faculty Affiliate}, \orgname{University of Toronto}, \orgaddress{\state{Toronto}, \country{Canada}}}



\corres{*Jing Xu \email{kenny.xu@duke-nus.ed.sg}}

\abstract[Summary]{
Technological advancements have made it possible to deliver mobile health interventions to individuals. A novel framework that has emerged from such advancements is the just-in-time adaptive intervention (JITAI), which aims to suggest the right support to the individuals when their needs arise. 
The micro-randomized trial (MRT) design has been proposed recently to test the proximal effects of the components of these JITAIs. However, the extant MRT framework only considers components with a fixed number of categories added at the beginning of the study. 
We propose a more flexible MRT (FlexiMRT) design which allows addition of more categories to the components during the study. 
Note that the number and timing of the categories added during the study need to be fixed initially.
The proposed design is motivated by collaboration on the DIAMANTE study, which learns to deliver effective text messages to encourage physical activity among patients with diabetes and depression. We developed a new test statistic and the corresponding sample size calculator for the FlexiMRT using an approach similar to the generalized estimating equation for longitudinal data. Simulation studies were conducted to evaluate the sample size calculators and an R shiny application for the calculators was developed.
}


\maketitle

\section{Introduction}\label{Intro}
Mobile health (mHealth) is a term used to refer to the practice of medicine and health supported by mobile or wearables devices 
\cite{Sasan_2015} that are increasingly indispensable in our daily lives. 
It provides convenient support to various health domains including managing HIV infection \cite{Lewis_etal_2013}, increasing physical activity \cite{King_etal_2013}, supporting recovery from alcohol dependence \cite{Alessi_Petry_2013} and smoking cessation \cite{Free_etal_2013}.

Mobile technology can be used to deliver \textit{just-in-time adaptive interventions} (JITAIs), which aim to provide the right type or amount of support, 
at the right time \cite{Intille_2004,Patrick_etal_2008}, according to an individual's evolving internal and contextual state. 
Nahum-Shani \textit{et al} \cite{Nahum-Shani_etal_2018} bridge the gap between the growing technological capabilities for delivering JITAIs and the research on the development and evaluation of these interventions.  

The Micro-randomized trial (MRT) design \cite{Klasnja_etal_2015} has been proposed for testing the proximal effects of the intervention categories in JITAIs.
The corresponding sample size calculation has been derived by Liao \textit{et al} \cite{Liao_etal_2016}. 
In an MRT, there are numerous decision time points for each participant throughout the study period. At each decision time point, a participant is randomly assigned to one of the available intervention options.
There exist several research studies using the MRT design, for example, `HeartSteps' for promoting physical activity among sedentary people \cite{Klasnja_etal_2019}, `Sense2Stop' for managing stress in newly abstinent smokers \cite{Liao_etal_2018}, 
`DIAMANTE' \cite{Aguilera_etal_2020} for promoting physical activity among co-morbid diabetes and depression patients, 
`StayWell' \cite{Figueroa_etal_2020} for managing people's mental wellness during COVID-19 pandemic period, 
and so on.  

In the existing MRT design, the categories of an intervention component are predetermined and fixed during the trial. 
With the goal to expand its scope, similar to adding new treatment arms during the study in platform trials (e.g., \cite{Ventz_etal_2017} and \cite{Lee_etal_2019}), we propose a novel flexible version of the MRT (``FlexiMRT") design that allows newly specified intervention categories to be added not only initially but also later in the study. 
Unlike the platform trials, however, the timings of adding new intervention categories and the numbers of the intervention categories to be added at these times are pre-determined in FlexiMRTs.
Going beyond prespecified categories is often needed to speed up research. 
For example, it enables using in-trial qualitative participant feedback and suggestions, to design and include new conditions, which can then be tested quantitatively. 
No matter how extensive, predeployment thinking or theory about what conditions to test can often miss better ideas, which only arise in the real-world context of diverse participants experiencing the intervention. 
For example, beyond health domains, rapid iterative cycles of design \cite{Mei_etal_2021} is key to success in disciplines like human computer interaction and software product testing. 
Our methods support faster cycles of experimentation targeted at ecologically valid questions.
In addition, novel techniques like crowdsourcing \cite{DeVries_etal_2016} enable applications in fields from social media to health to education where ideas are emerging after deployment. 
For example, Williams \textit{et al} \cite{Williams_etal_2016} shows the value of having new explanations of a concept continually added by people learning a topic, once they can be tested out quickly to assess efficacy. 
More broadly, applications of experimentation for successful apps, websites and products rely on cycles of continual testing and improvement \cite{Koning_etal_2022}. 
Unlike the traditional long delay from trial to deployment, an app like Facebook or Google search will perpetually be having product teams add and modify conditions for testing \cite{Fabijan_etal_2018}, on the order of thousands of experimental changes/deployments per day, since constantly changing product needs and context require more dynamic experimentation techniques. 
Our work aims to provide sounder methodological basis for such experiments in mHealth and beyond, to bring innovative techniques to bear on scientifically supported improvements for health interventions to help people.

In our proposed FlexiMRT design, for each intervention component, instead of estimating one proximal effect (i.e., the ``pooled'' intervention categories versus the control category), the proximal effects of individual intervention categories versus no intervention (control) can be estimated. 
The individual effect sizes will be of interest if they are expected to be different in magnitudes or directions.
Novel test statistics are derived to detect the proximal effects from the FlexiMRT data. 
We derive the corresponding sample size calculators so that the trials can be sized to either detect the proximal effect at a nominal power and a nominal type-I error rate, or to estimate the proximal effect within a certain precision or margin of error at a nominal confidence level.
The latter approach is useful in a pilot study (e.g., the DIAMANTE students study by Figueroa \textit{et al} \cite{Figueroa_etal_2020b}
) scenario, when prior information on the proximal effects may be limited. 

This work is motivated by our collaboration on the Diabetes and Mental Health Adaptive Notification Tracking and Evaluation (DIAMANTE) study \cite{Aguilera_etal_2020}. The ``DIAMANTE'' app sends multiple messages to encourage physical activity among co-morbid diabetes and depression patients from low-income and ethnic minority backgrounds served in the San Francisco Health Network. 

We summarize the three main contributions of the paper. 
First, we propose a novel FlexiMRT design, which allows for intervention categories to be added later in the study. 
Second, we develop two associated sample size calculation methods (one based on power and the other on precision) for the proposed design.
Finally, based on the developed methodology, we create free online sample size calculators using R shiny (https://kennyxu.shinyapps.io/FlexiMRT-SS/) to facilitate wide dissemination.
The relevant R functions are available in Github (https://github.com/Kenny-Jing-Xu/FlexiMRT-SS).

The rest of the paper is organized as follows. Section \ref{s:analysissamplesize} describes the proposed FlexiMRT design with the corresponding statistical analysis and sample size calculation approach. 
Section \ref{s:realexample} demonstrates an application of the proposed sample size calculators based on 
an adapted student version of the DIAMANTE study. Section \ref{s:sim} investigates the performance of the sample size calculators through simulation studies. 
The paper ends with a discussion in Section \ref{s:disc}. The detailed derivations and the implementation of the proposed sample size calculators are deferred to the Supplementary Material.

\section{FlexiMRT Design: Statistical Model Estimation and Sample Size Calculation}\label{s:analysissamplesize}

\subsection{MRT Design}\label{s:MRT}
MRT is a cutting-edge trial design suitable for time-varying, sequential, multi-component interventions, akin to a sequential full-factorial design. 
Within the study period, at decision time $t$, on day $d$, participant $i$ is randomized to an intervention component denoted by $A_{idt}$, where $i$=$1$, $\ldots$, $N$, $d$=$1$, $\ldots$, $D$ and $t$=$1$, $\ldots$, $T$.
For example, $A_{idt}=1$ if an intervention category is delivered with probability $\pi_{dt}$, otherwise  $A_{idt}=0$ if a control category is delivered.
Though the extant MRT design \cite{Liao_etal_2016} mainly focuses on components with binary categories only, we consider multiple categories in the current paper.
Let $Y_{idt}$ denote the proximal outcome of participant $i$ measured following time point $t$ on day $d$.

A participant may not be in a position to receive an intervention at a decision point during the study. 
For example, it is not safe for an mHealth app to deliver an intervention to a participant who is driving. 
A participant is considered unavailable for randomization at such decision points, therefore no intervention or control is delivered automatically. 
Let $I_{idt}$ denote the availability indicator, i.e., $I_{idt} =1$ if participant $i$ is available at decision point $t$ on day $d$ and $I_{idt} =0$ otherwise. 
Note that $A_{idt}$=$0$ if $I_{idt}$=$0$ and the randomization probability is denoted by $\pi_{dt}$ = $P(A_{idt} =1 \mid I_{idt} =1)$.

MRTs consider the longitudinal data, i.e., $\boldsymbol O_i$ =($\boldsymbol S_i^{\top}$, $I_{i11}$, $A_{i11}$, $Y_{i11}$,$\ldots$, $I_{iDT}$, $A_{iDT}$, $Y_{iDT}$) for the observations of participant $i$, where $\boldsymbol S_i$ denotes the baseline covariates vector. 
We assume that $\boldsymbol O_i$,  $i=1,\ldots,N$ are independent and identically distributed.
Note that the sample size calculation focuses on detecting the proximal effect of intervention. 

\subsection{DIAMANTE Study}\label{s:MLMRT}
\begin{figure}
\begin{center}
\begin{tabular}{c}
 \includegraphics[height=7cm, width=12cm]{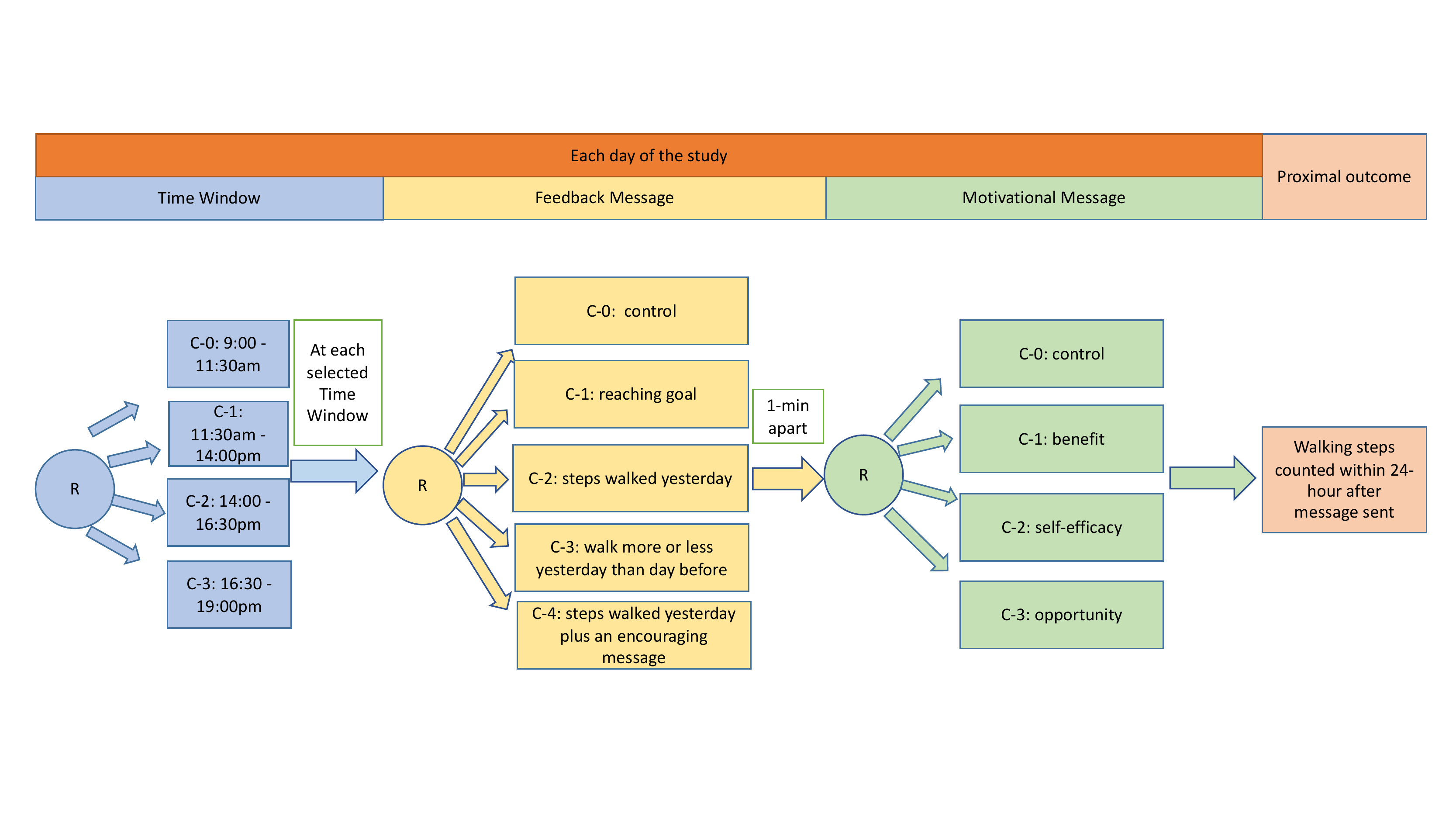}
\end{tabular}
\end{center}
\caption{
MRT Design for DIAMANTE study,
R=Randomization and m=Category. 
}\label{fig: MLMRT_DIAMANTE}
\end{figure}

The FlexiMRT design is motivated by the DIAMANTE study that employs an
MRT design with multi-category components and is summarized in Figure \ref{fig: MLMRT_DIAMANTE}.
This is a six-month study with one decision time point per day.
On each day, each available participant is randomized to one of the categories of each of three 
components, namely, {\em Time Window} (4 categories), {\em Feedback Message} (5 categories), and {\em Motivational Message} (4 categories). 
Thus, the DIAMANTE study allocates interventions according to a $4 \times 5 \times 4$ factorial design each day.  
The two different messages are sent one minute apart. 

\textbf{The Time Window component} ($\boldsymbol A^T$)
specifies when to send the messages, i.e. category-0 (9:00 am - 11:30 am), category-1 (11:30 am - 14:00 pm), category-2 (14:00 pm - 16:30 pm) and category-3 (16:30 pm - 19:00 pm). 

\textbf{The Feedback Message component} ($\boldsymbol A^F$)
has a reference category-0 (i.e. no message) and four intervention categories,  i.e. category-1 (reaching goal), category-2 (steps walked yesterday), category-3 (walked more or less yesterday than the day before), category-4 (steps walked yesterday plus a positive/negative message).

\textbf{The Motivational Message component\textbf} ($\boldsymbol A^M$)
has a reference category-0 (no message) and three intervention categories, i.e. category-1 (benefit), category-2 (self-efficacy) and category-3 (opportunity). 

\textbf{The proximal outcome} ($Y$) 
is the daily steps change,
i.e., the step-count on the calendar day when intervention messages were sent minus the step-count on the previous calendar day, measured by the participants' phone pedometer, whose formula is defined in Section \ref{s:analysis}.

\textbf{The availability indicator} ($I$) at a given decision time codes one if a participant is available for intervention at that decision time and zero otherwise.
The DIAMANTE study assumes 
$100\%$ availability.
This is reasonable because the proximal outcome is the change in daily steps following the messages, thus allowing participants sufficient time to respond to messages at the randomly selected time windows.

The current phase of the DIAMANTE trial is a conventional MRT where all the intervention categories are introduced upfront. 
At the macro level, this study involves a randomized control trial with 3 groups, namely, the uniform randomization (UR) group, the adaptive learning (AL) group and the control group, where the MRT design is embedded within the UR and AL groups.
The participants of the UR group receive different messages at different times with equal probability while the participants of the AL group receive them with probabilities learned adaptively by a reinforcement learning algorithm aimed at maximizing the proximal effects. 
Though both the UR and AL groups can be modified into a FlexiMRT design, the proposed sample size calculators can be only used for the UR group.

For the statistical model and sample size calculators proposed through Sections \ref{s:analysis} to \ref{powersamplesize},
we consider the long format of longitudinal datasets.
The observation on participant $i$ on day $d$ at decision time point $t$ is
\begin{equation*}
\{I_{idt}, A^{T}_{1idt}, A^{T}_{2idt}, A^{T}_{3idt},  A^{F}_{1idt}, A^{F}_{2idt}, A^{F}_{3idt},  A^{F}_{4idt}, A^{M}_{1idt}, A^{M}_{2idt}, A^{M}_{3idt}, Y_{idt}\},
\end{equation*}
where $\{A^{T}_{1}, A^{T}_{2}, A^{T}_{3}\}$, $\{A^{F}_{1}, A^{F}_{2}, A^{F}_{3}, A^{F}_{4}\}$ and $\{A^{M}_{1}, A^{M}_{2}, A^{M}_{3}\}$ represent the intervention category indicators of the Time Window, Feedback Message and Motivational Message components respectively.

\subsection{Statistical Model}\label{s:analysis}
In this section, we first extend the statistical model for MRT \cite{Klasnja_etal_2015} proposed by Liao \textit{et al} \cite{Liao_etal_2016} to accommodate multi-category intervention components. 
Instead of estimating the proximal effect between two categories (active versus control), we estimate such effects for multiple active categories with reference to control.  
The multi-category approach can be used to recognize whether the differences in effect sizes among the intervention categories are in magnitudes only (see Section \ref{s:realexample}) or if opposite direction of signs exist.
A regression model considering the proximal effect of a single component can be defined as below. We assume one decision time point per day, i.e. $T=1$ for participant $i$ on day $d$, with the proximal outcome denoted by $Y_{id}$. 
Note that the decision is to allocate a participant to one of the intervention categories. The allocation only depends on the randomization probabilities at the same time point, and does not depend on the outcome and randomization probabilities at the previous time points. 

We define the proximal effect size of the message category $m$ (versus the control category $0$) of a particular component on day $d$, denoted by a function $b_m(d; \boldsymbol\beta_m)$, as
\begin{equation}\label{Zbeta_p}
b_m(d; \boldsymbol\beta_m)=Z_{md}^{\top}\boldsymbol\beta_m = \left[ 1, d-1, \ldots, ( d - 1 )^{p_m-1} \right]\boldsymbol\beta_m,
\end{equation}
where $m=1,\ldots,M$, $d=1,\ldots,D$ (the study period in days) and parameter $\boldsymbol\beta_m=(\beta_{m1},\ldots,\beta_{mp_m})^{\top}$. 

Note that $\boldsymbol Z_{md}^{\top}$ in equation (\ref{Zbeta_p}) corresponds to one decision time point per day as in DIAMANTE, which can be generalized by involving the number of decision time points $T$ per day. 
We define the proximal effect size of category $m$ at time point $t$ of day $d$ from a particular component, denoted by $b_m(d, t; \boldsymbol\beta_m)$, as
\begin{equation*}
b_m(d, t; \boldsymbol\beta_m)=Z_{mdt}^{\top}\boldsymbol\beta_m = \left( 1, \left[ \dfrac{( d - 1 )T+t-1}{T} \right], \ldots, \left[  \dfrac{( d - 1 )T+t-1}{T} \right]^{p_m-1} \right)\boldsymbol\beta_m,
\end{equation*}
where $t=1, \ldots, T$. 
The proximal effects of different categories on different days can vary. 
They may follow the constant, linear, or quadratic trends, corresponding to $p_m=$1, 2 or 3 respectively.
Alternatively, 
the proximal effect trend for category-$m$ can also be described as having a combination of the linear and constant trends, where it increases or decreases linearly until a turning point on day $d_{\text{turn}}^m$ and plateaus afterwards, 
as demonstrated in the StayWell study results \cite{Figueroa_etal_2020}. 
We call it the ``linear-plateau" trend. In this case, we can define the proximal effect size using a linear spline, i.e.,
\begin{equation}\label{Zbeta_lspline}
b_m(d, t; \boldsymbol\beta_m)=Z_{mdt}^{\top}\boldsymbol\beta_m =  \left( 1, \left[ \dfrac{ \text{min}\left[d_{\text{turn}}^m -1, d - 1 \right] T+t-1}{T} \right] \right)\boldsymbol\beta_m,
\end{equation}
where we have $\boldsymbol\beta_m^{\top}$=($\beta_{m1}, \beta_{m2})$.

Next, we consider the scenario where new message categories are added during the study. 
Let $M_0$ be the number of categories introduced initially on day $d_{0}$ (e.g., $d=1$), $M_1$ be the  number of message categories added on day $d_{1}$ (first adding day after the beginning of the study) and so on, and finally $M_k$ be the number of categories added at the last adding day $d_k$, where we have $d_{0}\leq 1<d_{1}<\ldots <d_{k}\leq D$. 
Therefore, the total number of message categories $M=\sum_{j=0}^{k}M_j$. 
Note that for the experimental design and sample size calculation purposes, $M_j$ and $d_j$, where $j=1,\ldots,k$, are pre-determined. 
We assume that the participants have the same length of time until each new category is added. This ensures that all added messages have the same duration for each participant to estimate their proximal effects. 
Note that the effect size of each of new message category is undefined before their adding days. 
Suppose $d=d_1$, $m=M_0+1$, $T=1$ and $p_m=2$, then the proximal effect of category $M_0+1$ on day $d_1$ can be computed by $b_{M_0+1}(d_1; \boldsymbol\beta_{M_0+1})$=$\beta_{ ( M_0 + 1 ) 1 } + \beta_{ ( M_0 + 1 ) 2 }(d_1-1)$ based on equation (\ref{Zbeta_p}), where $b_{M_0+1}(d_1; \boldsymbol\beta_{M_0+1})$ is undefined before day $d_1$.
Given that participant $i$ is available for randomization on day $d$, we denote the message categories by $\boldsymbol A_{id}$=($A_{i1d}$,$\ldots$,$A_{iM_0d}$,$\ldots$,$A_{i( \sum_{j=0}^{k-1}M_j+1 )d}$,$\ldots$,$A_{i( \sum_{j=0}^{k}M_j )d}$)$^{\top}$ and assume that they follow a multinomial distribution, i.e. \textit{Multinomial}(1-$\sum_m\pi_{md}$, $\pi_{1d}$,$\ldots$,$\pi_{( \sum_{j=0}^{k}M_j  )d}$), where $\pi_{(\cdot)}$ denotes the randomization probability corresponding to $A_{i(\cdot)}$. Note that before the first adding day (i.e. $d < d_1$), the message categories $M_0 + 1$ to $\sum_{j=0}^{k}M_j$ are not available. Thus, the category indicators $A_{ i ( M_0 + 1 ) d }$,$\ldots$,
$A_{i( \sum_{j=0}^{k}M_j )d}$ and their corresponding randomization probabilities $\pi_{ i ( M_0 + 1 ) d }$,$\ldots$, $\pi_{i( \sum_{j=0}^{k}M_j )d}$ have zero values. 
Therefore, to allow for the flexible addition of new intervention categories during the trial, we further remove the restriction on fixed allocation in the model.
We then denote the proximal effects of all the message categories on day $d$ to be 
\begin{align*}
\boldsymbol Z_{1 d}^{\top}\boldsymbol\beta_1, \ldots, \boldsymbol Z_{M_0 d}^{\top}\boldsymbol\beta_{M_0},\ldots, \boldsymbol Z_{\sum_{j=0}^{k-1}M_j+1 d}^{\top}\boldsymbol\beta_{\sum_{j=0}^{k-1}M_j+1}, \ldots, \boldsymbol Z_{\sum_{j=0}^{k}M_j d}^{\top}\boldsymbol\beta_{\sum_{j=0}^{k}M_j}.
\end{align*}
The working model can be written as 
\begin{align*}
Y_{id}&=\boldsymbol B_{d}^{\top}\boldsymbol\alpha\\
&+
(A_{i1d}-\pi_{1d})\boldsymbol Z_{1 d}^{\top}\boldsymbol\beta_1+ \cdots +(A_{iM_{0}d}-\pi_{M_{0}d})\boldsymbol Z_{M_0 d}^{\top}\boldsymbol\beta_{M_0} \\
&+
(A_{i( M_0 + 1 )d}-\pi_{( M_0 + 1 )d})\boldsymbol Z_{( M_0 + 1 ) d}^{\top}\boldsymbol\beta_{ ( M_0 + 1 ) }+ \cdots \\
&+(A_{i( M_{0} + M_{1} ) d}-\pi_{( M_{0} + M_{1} )d})\boldsymbol Z_{( M_{0} + M_{1} ) d}^{\top}\boldsymbol\beta_{( M_{0} + M_{1} )} \\
&\vdots\\
&+
(A_{i( \sum_{j=0}^{k-1} M_j +1) d}-\pi_{( \sum_{j=0}^{k-1}M_j + 1 )d})\boldsymbol Z_{( \sum_{j=0}^{k-1}M_j + 1 ) d}^{\top}\boldsymbol\beta_{ ( \sum_{j=0}^{k-1}M_j + 1 ) }+ \cdots \\
&+(A_{i( \sum_{j=0}^{k}M_j ) d}-\pi_{( \sum_{j=0}^{k}M_j )d})\boldsymbol Z_{( \sum_{j=0}^{k}M_j ) d}^{\top}\boldsymbol\beta_{( \sum_{j=0}^{k}M_j )} \\
&+\epsilon_{id}.
\end{align*}
$\boldsymbol B_{d}^{\top}\boldsymbol\alpha$ is a function of $d$ and covariates that are unaffected by intervention categories. 
For example, we can define a $(q-1)^{\text{th}}$-order function of $d$ for $B_{d}^{\top}\boldsymbol\alpha$ with parameters $\boldsymbol\alpha^{\top}$=$(\alpha_1, \ldots, \alpha_q)$ and $\boldsymbol B_{d}^{\top}$=$\left( 1, d - 1, \ldots, ( d - 1 )^{q-1} \right)$.
$\boldsymbol\beta_{m}^{\top}$=($\beta_{m1},\ldots,\beta_{mp_m})$ is the parameter vector of interest for the $m^{\text{th}}$ intervention category. 
$\epsilon_{id}$ is the error term,  
and $\boldsymbol\epsilon_{i}$=$(\epsilon_{i1},\ldots,\epsilon_{iD})^{\top}$ is assumed to follow a multivariate normal distribution with zero mean, variance $\sigma^2$ and correlation coefficient $\rho$, for $i=1,\ldots,N$. 
Note that the parameter $\rho$ can take nonzero values, meaning that observations from the same participants are not necessarily independent.
For each of the proximal effects $\boldsymbol Z_{m d}^{\top}\boldsymbol\beta_m$, $m=1,\ldots,M$, the intervention indicator $A_{imd}$ is centred by the randomization probability $\pi_{md}$,  as in \cite{Liao_etal_2016} 
and \cite{Boruvka_etal_2018}.
This centering procedure gives zero expected value for all $( A_{ i m d }-\pi_{ m d } )\boldsymbol Z_{ m d }^{\top}\boldsymbol\beta_m$, $m=1,\ldots,M$, and makes it easier to interpret the term $\boldsymbol B_{d}^{\top}\boldsymbol\alpha$ of the working model, i.e., $\boldsymbol B_{d}^{\top}\boldsymbol\alpha=E(Y_{id} \mid I_{id}=1)$, where $I_{id}$ is the availability indicator for participant $i$ on day $d$ with expectation $E(I_{id})=\tau_{d}$ \cite{Seewald_etal_2019}.

We define the model parameter for $\boldsymbol Y_i = (Y_{i1},\ldots,Y_{iD})^{\top}$, $i=1,\ldots,N$ by $\boldsymbol\theta$=($\boldsymbol\alpha^{\top}$, $\boldsymbol\beta^{\top}$)$^{\top}$, where $\boldsymbol\beta^{\top}$=($\boldsymbol\beta_1^{\top}$,$\ldots$,$\boldsymbol\beta_M^{\top}$). 
Following Liao \textit{et al} \cite{Liao_etal_2016} and Boruvka \textit{et al} \cite{Boruvka_etal_2018}, we derive the least squares (LS) estimator $\hat{\boldsymbol\theta}$ obtained by minimizing the squared error criterion, 
\begin{equation}
SEC(\boldsymbol\theta)=\dfrac{1}{N}\sum_{i=1}^{N}\sum_{d=1}^{D}I_{id}\left[ Y_{id}- \boldsymbol X_{id}^{\top}\boldsymbol\theta \right]^2,
\end{equation} 
where $N$ is the sample size
and $\boldsymbol X$ is the design matrix for the regression model of $\boldsymbol Y$. 
The LS estimator can be obtained by solving the equation
\begin{equation*}
\dfrac{\partial}{\partial\boldsymbol\theta}SEC(\boldsymbol\theta)=\dfrac{-2}{N}\sum_{i=1}^{N}\sum_{d=1}^{D}I_{id}\left[ Y_{id}-\boldsymbol X_{id}^{\top} \boldsymbol\theta \right]\boldsymbol X_{id}=0.
\end{equation*}
Therefore the LS estimator is 
\begin{equation}\label{theta_hat}
\hat{\boldsymbol\theta}=\left( \dfrac{1}{N}\sum_{i=1}^{N}\sum_{d=1}^{D}I_{id}\boldsymbol X_{id} \boldsymbol X_{id}^{\top} \right)^{-1}\dfrac{1}{N}\sum_{i=1}^{N}\sum_{d=1}^{D}I_{id}Y_{id}\boldsymbol X_{id},
\end{equation}
i.e. $\hat{\boldsymbol\theta}=(\hat{\boldsymbol\alpha}^{\top}, \hat{\boldsymbol\beta}^{\top})^{\top}$ with dimension $(q+\sum_{m=1}^M p_m) \times 1$, where $\hat{\boldsymbol\alpha}$ is a vector that includes the first $q$ elements of $\hat{\boldsymbol\theta}$ while $\hat{\boldsymbol\beta}$ includes the rest of the $\sum_{m=1}^M p_m$ elements of $\hat{\boldsymbol\theta}$.

\subsection{Test Statistics}\label{test}
This section proposes the hypothesis tests and derives the test statistics for the model parameters estimated in Section \ref{s:analysis}. 
We propose a null hypothesis that the proximal effect sizes of all intervention categories are zeros. In other words, the intervention components do not have any effects on the proximal outcome, i.e.,  
\begin{equation*}
H_{0}: b_m(d; \boldsymbol\beta_m)=0, \text{for all $m$ and $d$, or } \boldsymbol\beta=\boldsymbol 0,
\end{equation*} 
where $\boldsymbol 0$ represents a zero vector with $\sum_{m=1}^M p_m$ elements. An alternative hypothesis is that the proximal effect sizes of intervention categories are not all zeros. In other words, the intervention components have some effects on the proximal outcome, i.e., 
\begin{equation*}
H_{1}: b_m(d; \boldsymbol\beta_m)\neq 0, \text{for some $m$ and $d$, or }\boldsymbol\beta\neq\boldsymbol 0, 
\end{equation*}
where $m$ indicates the $m^{\text{th}}$ message category, i.e. $m=1,\ldots,M_0,\ldots,M_0+M_1,\ldots,\sum_{j=0}^{k}M_j=M$, $\boldsymbol\beta_m=(\beta_{m1},\ldots, \beta_{mp_m})^{\top}$; $d$ is the day during the study period, i.e. $d$=$1,\ldots,D$; and $\boldsymbol Z_{md}$ is defined in equation (\ref{Zbeta_p}). We have the control category at $m=0$. 
Note that the above hypothesis test is conducted over the entire study period, not at each decision time point separately.

In order to derive the test statistic distributions, the following assumptions are required. 
\begin{enumerate}
\item[1.] Let $\boldsymbol\Theta$ be the parameter space for $\boldsymbol\theta$, where $\boldsymbol\Theta$ is a compact subset of $R^{q+\sum_{m=1}^{M}p_m}$. 
\item[2.] $E(SEC(\boldsymbol\theta))$ exists and has a unique minimum value at $\tilde{\boldsymbol\theta}\in\boldsymbol\Theta$. 
\item[3.] $SEC(\boldsymbol\theta)$  is continuous, bounded, and differentiable in the neighbourhood of $\tilde{\boldsymbol\theta}$.
\item[4.] The matrix $\sum_{d=1}^{D}E\left( I_{id} \boldsymbol X_{id} \boldsymbol X_{id}^{\top} \right)$ in equation (\ref{theta_tilde_aa}) is invertible.
\end{enumerate}
First we present the following lemma about consistency and asymptotic normality for the least squares estimator $\hat{\boldsymbol\theta}$. 
The proof is deferred to the Supplementary Material.
\begin{lemma}\label{theo1}
The least squares estimator $\hat{\boldsymbol\theta}$ is a consistent estimator of $\tilde{\boldsymbol\theta}$. Under standard moment conditions and assumptions 1 - 4, we have $\sqrt{N}(\hat{\boldsymbol\theta}-\tilde{\boldsymbol\theta})\rightarrow \text{Normal}(0, \boldsymbol\Sigma_{\boldsymbol\theta})$.
\end{lemma}

When the sample size $N$ is large, the sample mean within the estimator (\ref{theta_hat}) is replaced by its expectation, i.e.,  
\begin{equation}\label{theta_tilde_aa}
\tilde{\boldsymbol\theta}=\left[\sum_{d=1}^{D}E\left( I_{id} \boldsymbol X_{id} \boldsymbol X_{id}^{\top} \right)\right]^{-1} \sum_{d=1}^{D}E\left( I_{id}Y_{id}\boldsymbol X_{id} \right),
\end{equation}
i.e., $\tilde{\boldsymbol\theta}$=$(\tilde{\boldsymbol\alpha}^{\top}, \tilde{\boldsymbol\beta}^{\top})^{\top}$. In other words, it turns out that $\hat{\boldsymbol\theta}\rightarrow\tilde{\boldsymbol\theta}$ when $N\rightarrow\infty$. More details about equation (\ref{theta_tilde_aa}) are covered in the Supplementary Material. 
The asymptotic covariance matrix  
$\boldsymbol\Sigma_{\boldsymbol\theta}$ is defined by
\begin{equation}\label{Sigma_theta}
\boldsymbol\Sigma_{\boldsymbol\theta} = \left[ \sum_{d=1}^{D}E( I_{id}\boldsymbol X_{id} \boldsymbol X_{id}^{\top}) \right]^{-1} 
E\left(\sum_{d=1}^{D} I_{id}\tilde{\epsilon}_{id}\boldsymbol X_{id}  \sum_{d=1}^{D} I_{id}\tilde{\epsilon}_{id}\boldsymbol X_{id}^{\top} \right) \left[ \sum_{d=1}^{D}E( I_{id}\boldsymbol X_{id} \boldsymbol X_{id}^{\top}) \right]^{-1},
\end{equation} 
where $E\left(\sum_{d=1}^{D} I_{id}\tilde{\epsilon}_{id}\boldsymbol X_{id}  \sum_{d=1}^{D} I_{id}\tilde{\epsilon}_{id}\boldsymbol X_{id}^{\top}  \right)$ is defined in the Supplementary Material. Thus the asymptotic distribution of $\hat{\boldsymbol\beta}$ converges to normal, i.e.
$\sqrt{N}(\hat{\boldsymbol\beta}-\tilde{\boldsymbol\beta})\rightarrow$ Normal(0, $\boldsymbol\Sigma_{\boldsymbol\beta}$) with covariance matrix $\boldsymbol\Sigma_{\boldsymbol\beta}$; see the Supplementary Material for its derivation. 
The asymptotic covariance matrix can be expressed as $\boldsymbol\Sigma_{\boldsymbol\beta}$=$\boldsymbol Q^{-1} \boldsymbol W \boldsymbol Q^{-1}$ with square matrices $\boldsymbol Q$ and $\boldsymbol W$ that are the lower right $\sum_{m=1}^{M}p_m\times\sum_{m=1}^{M}p_m$ blocks of  
$\sum_{d=1}^{D}E( I_{id}\boldsymbol X_{id} \boldsymbol X_{id}^{\top})$ and 
$E\left(\sum_{d=1}^{D} I_{id}\tilde{\epsilon}_{id}\boldsymbol X_{id}  \sum_{d=1}^{D} I_{id}\tilde{\epsilon}_{id}\boldsymbol X_{id}^{\top} \right) $, 
respectively.

As $\hat{\boldsymbol\beta}$ follows a normal distribution when the null hypothesis is true and $N$ is large, in a similar fashion as in Tu \textit{et al} \cite{Tu_etal_2004}, the test statistic function $C_N(\cdot)$, i.e., $C_N(\hat{\boldsymbol\beta})$=$N\hat{\boldsymbol\beta}^{\top}\boldsymbol\Sigma_{\boldsymbol\beta}^{-1}\hat{\boldsymbol\beta}$ follows a $\chi^2$ distribution with a degrees of freedom (df) of $\sum_{m=1}^{M} p_m$ (i.e., the length of $\boldsymbol\beta$).
For example, assuming a quadratic trend (i.e., $p_m=3$ for all $m = 1, \ldots, M$), the df of $N\hat{\boldsymbol\beta}^{\top}\boldsymbol\Sigma_{\boldsymbol\beta}^{-1}\hat{\boldsymbol\beta}$ for the motivation component ($M=3$) of the DIAMANTE study is $9$.
Unlike the Wald statistic, the Chi-square statistic allows the hypothesis to involve not only a single parameter but also a vector parameter.
We have $\boldsymbol\delta=\boldsymbol\beta/\bar{\sigma}$, where $\bar{\sigma}^2$ is the average of the residual variance over all the decision time points, i.e. $\bar{\sigma}^2$=$\sum_{d=1}^{D}\text{Var}(\epsilon_{id})/D$, 
therefore, $C_N(\hat{\boldsymbol\beta})$ = $C_N(\hat{\boldsymbol\delta})$ =$N\hat{\boldsymbol\delta}^{\top}(\boldsymbol\Sigma_{\boldsymbol\beta} /  \bar{\sigma}^2)^{-1}\hat{\boldsymbol\delta}$.
Tu \textit{et al} 
proposes a sample size calculation method based on power under the GEE approach for longitudinal data. 
Leveraging on this, we define the power function as follows.
If $H_0$: $\boldsymbol\beta=0$ is true, then the type-I error rate is defined by
\begin{equation}\label{N_chi_alpha}
\text{Pr}\left(X_{ \sum_{m=1}^{M}p_m } > \chi^2_{ \sum_{m=1}^{M}p_m, \alpha}\right)=\alpha,
\end{equation}
where $X_{ \sum_{m=1}^{M}p_m }$ presents a random variable following a central Chi-squared distribution with degrees of freedom $\sum_{m=1}^{M}p_m$ while $\chi^2_{ \sum_{m=1}^{M}p_m, \alpha}$ represents its $1-\alpha$ quantile. 
We reject $H_0$ at level $\alpha$ if $X_{ \sum_{m=1}^{M}p_m } > \chi^2_{ \sum_{m=1}^{M}p_m, \alpha}$.
If $H_1$: $\boldsymbol\beta=\tilde{\boldsymbol\beta}\neq 0$ is true, then
\begin{equation}\label{N_chi_P}
\text{Pr}\left(X_{ \sum_{m=1}^{M}p_m, C_N(\tilde{\boldsymbol\delta} )} > \chi^2_{\sum_{m=1}^{M}p_m, \alpha}\right)=\text{Power},
\end{equation}
where $X_{ \sum_{m=1}^{M}p_m,  C_N(\tilde{\boldsymbol\delta})}$ represents a random variable following a Chi-squared distribution with degrees of freedom $\sum_{m=1}^{M}p_m$.
Note that \cite{Tu_etal_2004} does not define the distribution of $C_N(\hat{\boldsymbol\delta})$ for a small sample size.

When $N$ is small, $\boldsymbol\Sigma_{\boldsymbol\beta}$ is replaced by its sample estimate $\hat{\boldsymbol\Sigma}_{\boldsymbol\beta}$, which is derived by Mancl and DeRouen \cite{Mancl_DeRouen_2001}, 
and the test statistic follows Hotelling's $T^2$ distribution; see, e.g., \cite{Hotelling_1931}, and \cite{Li_Redden_2015}. 
We define the small-sample estimator by $\hat{\boldsymbol\Sigma}_{\boldsymbol\beta}$=$\hat{\boldsymbol Q}^{-1} \hat{\boldsymbol W} \hat{\boldsymbol Q}^{-1}$. Let
\begin{equation}\label{e_hat_id}
\hat{e}_{id}=Y_{id} - \boldsymbol X_{id}^{\top}\hat{\boldsymbol\theta},
\end{equation}
\begin{equation}\label{e_hat_i}
\hat{\boldsymbol e}_{i}^{\top}=(\hat{e}_{i1}, \ldots, \hat{e}_{iD}),
\end{equation}
\begin{equation}\label{X_i}
\boldsymbol X_{i}^{\top}=
\begin{bmatrix}
\boldsymbol X_{i1}^{\top}I_{i1}\\
\vdots \\
\boldsymbol X_{iD}^{\top}I_{iD}
\end{bmatrix}_{D\times (q+ \sum_{m=1}^{M}p_m)},
\end{equation}
\begin{equation}\label{H_i}
\boldsymbol H_{i}=\boldsymbol X_{i}^{\top}[\sum_{i=1}^N \boldsymbol X_{i} \boldsymbol X_{i}^{\top}]^{-1} \boldsymbol X_{i}.
\end{equation}
The matrix $\hat{\boldsymbol Q}^{-1}$ is given by the lower right $\sum_{m=1}^{M}p_m\times \sum_{m=1}^{M}p_m$ block of $[\sum_{i=1}^N \boldsymbol X_{i} \boldsymbol X_{i}^{\top}/N]^{-1}$; the matrix $\hat{\boldsymbol W}$ is given by the lower right $\sum_{m=1}^{M}p_m\times \sum_{m=1}^{M}p_m$ block of $[\sum_{i=1}^N \boldsymbol X_{i} (\boldsymbol I_{D\times D} - \boldsymbol H_i)^{-1} \hat{\boldsymbol e}_{i}\hat{\boldsymbol e}_{i}^{\top} (\boldsymbol I_{D\times D} - \boldsymbol H_i)^{-1} \boldsymbol X_{i}^{\top}]/N$, where $\boldsymbol I_{D\times D}$ is the identity matrix with dimension $D\times D$. 

Liao \textit{et al} \cite{Liao_etal_2016} suggests that the test statistic follows a Hotelling's $T^2$ distribution with dimension $\sum_{m=1}^{M}p_m$ and degrees of freedom $N-q-1$, i.e.,
\begin{align*}
\hat{C}_N(\hat{\boldsymbol\delta})=N\hat{\boldsymbol\beta}^{\top}\hat{\boldsymbol\Sigma}_{\boldsymbol\beta}^{-1}\hat{\boldsymbol\beta}\sim T^2_{ \sum_{m=1}^{M}p_m, N-q-1}=\dfrac{ \sum_{m=1}^{M}p_m(N-q-1)}{N-q- \sum_{m=1}^{M}p_m} F_{ \sum_{m=1}^{M}p_m, N-q- \sum_{m=1}^{M}p_m}.
\end{align*} 
Thus,
\begin{align*}
\dfrac{N-q- \sum_{m=1}^{M}p_m}{ \sum_{m=1}^{M}p_m(N-q-1)} \hat{C}_N(\hat{\boldsymbol\delta})\sim F_{ \sum_{m=1}^{M}p_m, N-q- \sum_{m=1}^{M}p_m}
\end{align*}
if $H_0$: $\boldsymbol\beta=0$ is true, and the type-I error rate ($\alpha$) is defined by
\begin{equation}\label{N_F_N_q_1_alpha}
\text{Pr}\left(F_{ \sum_{m=1}^{M}p_m, N-q-\sum_{m=1}^{M}p_m} > f_{\sum_{m=1}^{M}p_m, N-q-\sum_{m=1}^{M}p_m, \alpha}\right)=\alpha,
\end{equation}
where $F_{ \sum_{m=1}^{M}p_m,  N-q-\sum_{m=1}^{M}p_m}$ represents a random variable following a central $F$ distribution with degrees of freedom $\sum_{m=1}^{M}p_m$ and $N-q-\sum_{m=1}^{M}p_m$ while $f_{\sum_{m=1}^{M}p_m, N-q-\sum_{m=1}^{M}p_m, \alpha}$ represents its $1-\alpha$ quantile.
We reject $H_0$ at level $\alpha$ if $F_{ \sum_{m=1}^{M}p_m, N-q-\sum_{m=1}^{M}p_m} > f_{\sum_{m=1}^{M}p_m, N-q-\sum_{m=1}^{M}p_m, \alpha}$.
If $H_1$: $\boldsymbol\beta=\tilde{\boldsymbol\beta}\neq 0$ is true, then 
\begin{align*}
\dfrac{N-q-\sum_{m=1}^{M}p_m}{ \sum_{m=1}^{M}p_m(N-q-1)} \hat{C}_N(\hat{\boldsymbol\delta})\sim F_{ \sum_{m=1}^{M}p_m, N-q- \sum_{m=1}^{M}p_m, \tilde{C}_N(\tilde{\boldsymbol\delta})},
\end{align*}
i.e. a non-central $F$ distribution with non-centrality parameter 
$\tilde{C}_N(\tilde{\boldsymbol\delta})$ = $N\tilde{\boldsymbol\delta}^{\top}(\tilde{\boldsymbol\Sigma}_{\boldsymbol\beta} /  \bar{\sigma}^2)^{-1}\tilde{\boldsymbol\delta}$, where $\tilde{\boldsymbol\Sigma}_{\boldsymbol\beta}$ is the sample estimate of $\boldsymbol\Sigma_{\boldsymbol\beta}$ when $N$ is large. 
Therefore
\begin{equation}\label{N_F_N_q_1_P}
\text{Pr}\left(F_{ \sum_{m=1}^{M}p_m, N-q- \sum_{m=1}^{M}p_m, \tilde{C}_N(\tilde{\boldsymbol\delta}) }> f_{ \sum_{m=1}^{M}p_m, N-q- \sum_{m=1}^{M}p_m, \alpha}\right)=\text{Power},
\end{equation}
where $F_{ \sum_{m=1}^{M}p_m,  N-q-\sum_{m=1}^{M}p_m, \tilde{C}_N(\tilde{\boldsymbol\delta}) }$ represents a random variable following a non-central $F$ distribution with degrees of freedom $\sum_{m=1}^{M}p_m$ and $N-q-\sum_{m=1}^{M}p_m$  and non-centrality parameter  $\tilde{C}_N(\tilde{\boldsymbol\delta}) $.
Note that Liao \textit{et al.} \cite{Liao_etal_2016} did not provide any mathematical proofs for the distribution of the test statistic. 

In this paper, we further suggest an alternative distribution for $\hat{C}_N(\hat{\boldsymbol\delta})$ (see Corollary \ref{coro1} below) and provide certain mathematical derivations in the Supplementary Material. 
\begin{corollary}\label{coro1}
According to Lemma \ref{theo1}, under a finite sample, the test statistic $\hat{C}_N(\hat{\boldsymbol\delta})=N\hat{\boldsymbol\beta}^{\top}\hat{\boldsymbol\Sigma}_{\boldsymbol\beta}^{-1}\hat{\boldsymbol\beta}$ follows a Hotelling's $T^2_{ \sum_{m=1}^{M}p_m, N}$ distribution.
\end{corollary}
This distribution can be defined by
 \begin{align*}
\hat{C}_N(\hat{\boldsymbol\delta})=N\hat{\boldsymbol\beta}^{\top}\hat{\boldsymbol\Sigma}_{\boldsymbol\beta}^{-1}\hat{\boldsymbol\beta}\sim T^2_{ \sum_{m=1}^{M}p_m, N}=\dfrac{ \sum_{m=1}^{M}p_m(N)}{N- \sum_{m=1}^{M}p_m +1 } F_{ \sum_{m=1}^{M}p_m, N- \sum_{m=1}^{M}p_m +1}
\end{align*} 
under $H_0$. 
The corresponding type-I error rate ($\alpha$) can be defined by
 \begin{equation}\label{N_F_N_alpha}
\text{Pr}\left(F_{ \sum_{m=1}^{M}p_m, N- \sum_{m=1}^{M}p_m +1} > f_{ \sum_{m=1}^{M}p_m, N- \sum_{m=1}^{M}p_m +1, \alpha}\right)=\alpha.
\end{equation}
We reject $H_0$ at level $\alpha$ if 
\begin{equation*}
F_{ \sum_{m=1}^{M}p_m, N- \sum_{m=1}^{M}p_m +1} > f_{ \sum_{m=1}^{M}p_m, N- \sum_{m=1}^{M}p_m +1, \alpha}.
\end{equation*}
In additon if $H_1$: $\boldsymbol\beta=\tilde{\boldsymbol\beta}\neq 0$ is true, then the power function can be defined by
 \begin{equation}\label{N_F_N_P}
\text{Pr}\left(F_{ \sum_{m=1}^{M}p_m, N- \sum_{m=1}^{M}p_m +1, \tilde{C}_N(\tilde{\boldsymbol\delta})} > f_{ \sum_{m=1}^{M}p_m, N- \sum_{m=1}^{M}p_m +1, \alpha}\right)=\text{Power}.
\end{equation} 
Simulation studies show that the proposed test statistic provides better power and coverage probability estimates than the test statistic of Liao \textit{et al.} \cite{Liao_etal_2016} when the sample size is small.

\subsection{Sample Size Calculation}\label{powersamplesize}
Here, we propose a power-based sample size formula for FlexiMRT.
This method requires a working knowledge of the standardized proximal effect sizes of intervention categories (i.e. $b_m(d; \boldsymbol \delta_m)=\boldsymbol Z_{md}^{\top}\boldsymbol \delta_m$, where $\boldsymbol \delta_m=\boldsymbol\beta_m/\bar{\sigma}$, for $m=1,\ldots,M$) as input parameters. 
Thus, given the desired power 
and the intended minimum detectable standardized $\boldsymbol\beta$ ($\boldsymbol\delta$), the sample size $N$ can be computed by solving for either equation (\ref{N_chi_P}), (\ref{N_F_N_q_1_P}),
or (\ref{N_F_N_P}), depending on the choice of the test statistic. 
For example, assuming a linear-plateau trend, $\boldsymbol\delta$ can be derived by the standardized initial $b_m(d_j; \boldsymbol\delta_m)$ and average $\dfrac{1}{D-(d_j-1)}\sum_{d=d_j}^{D}b_m(d;\boldsymbol\delta_m)$ proximal effect sizes and the turning point day $d_{\text{turn}}^{m}$, where the $m$-th category is added on day $d_{j}$, for $m=1,\ldots,M$ and $j=0, 1,\ldots,k$.
The calculated $N$ is therefore the minimum integer that gives the estimated power not lower than its nominal value.

\begin{figure}
\begin{center}
\begin{tabular}{ccc}
\includegraphics[height=4cm, width=4cm]{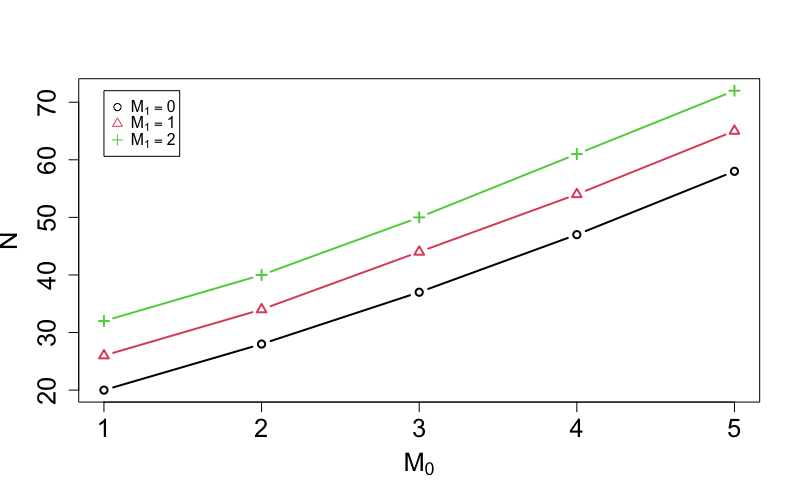} &
\includegraphics[height=4cm, width=4cm]{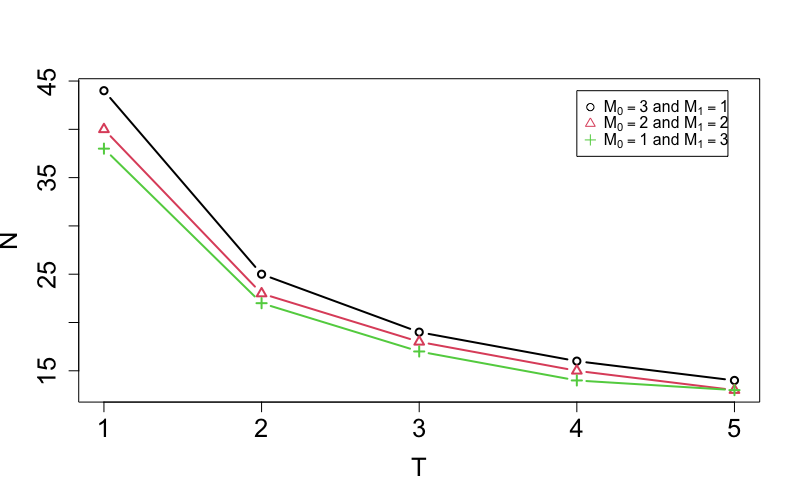} &  
\includegraphics[height=4cm, width=4cm]{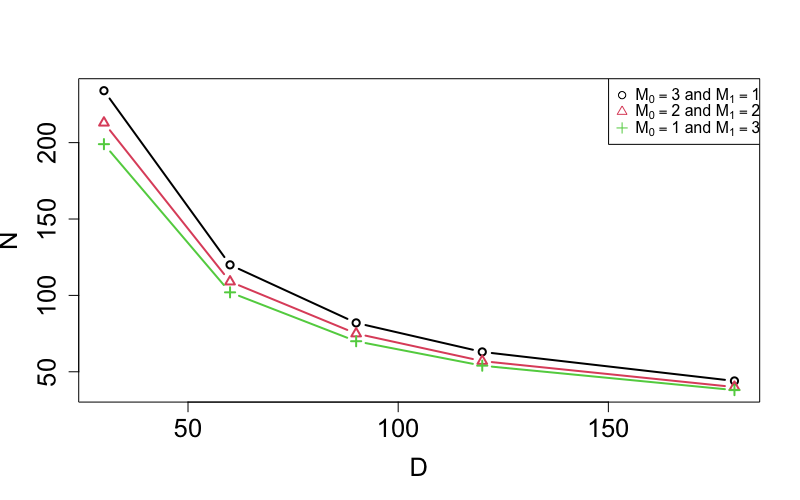} \\
a) & b) & c)
\end{tabular}
\end{center}
\caption{
Sample size $N$ versus: 
(a) initial intervention categories $M_0$ given different number of categories $M_1$ added  halfway through the study; (b) number of decision time points $T$ per day and c) study days $D$ given $M = 4$ and different $M_0$ and $M_1$ allocations. 
}
\label{fig: TvsN}
\end{figure}

In Figure \ref{fig: TvsN}, we illustrate the operating characteristics of the required sample size as a function of the design parameters, assuming constant trend proximal effect and using the Hotelling's $T^2$ test statistic (equation (\ref{N_F_N_q_1_P})).
In general, $N$ increases with the number of intervention categories $M$ in a particular component, but decreases with the number of decision time points and study days. Although the difference is slight, $N$ decreases when more categories are added later than earlier, i.e., fewer $M_0$ and more $M_1$ categories, for a fixed number of total categories (e.g. $M=M_0+M_1=4$). 

Figure \ref{fig: TvsN} a) also illustrates the benefit (i.e., avoiding the unnecessary increases in both the sample size and the study duration) of allowing additional categories to be added later in the trial. 
Take the StayWell study as an example (contact the authors of Figueroa \textit{et al} \cite{Figueroa_etal_2020} for relevant details), where two consecutive MRTs were conducted, the first with two categories and the latter with an additional category (three categories in total) at the beginning of each trial. 
Assuming all other design parameters are specified in Figure \ref{fig: TvsN} a), the total sample size needed for the two trials is $N$=$65$. In contrast, a FlexiMRT design would have suggested using the $M_0$ = $2$ and $M_1$ = $1$ variation, which only gives a sample size of $N$ = $34$.

Note the power-based method should only be used if the goal is to perform a hypothesis test, as is typical in a confirmatory study. However, due to the novelty and recency of MRT, prior information on the proximal effects of the individual categories are often limited. 
A pilot MRT to assess the feasibility and acceptability of the intervention categories may be necessary before conducting the full-scale study, similar to pilot studies for sequential multiple assignment randomized trials (SMARTs) \cite{Almirall_etal_2012}.  
To operationalize the sample size calculation for pilot MRT, we suggest the sample size to be calculated based on certain desired level of precision for the $\boldsymbol\beta$ estimates; 
see the Supplementary Material for the derivation,
Kelley \textit{et al} \cite{Kelley_etal_2003} or Maxwell \textit{et al} \cite{Maxwell_etal_2008} for discussions in the classical settings,
and Yan \textit{et al} \cite{Yan_etal_2020} for the rationale and benefit in the context of pilot SMARTs.    

\section{DIAMANTE Study with Student Population Example}\label{s:realexample}
A smaller study \cite{Figueroa_etal_2020b} was conducted where 93 students from the University of California, Berkeley used the DIAMANTE app (66 and 27 participants in the UR and AL groups respectively) for 45 days. A total of 44 decision time points and proximal outcome measures (corresponding to 45 days) were collected from each participant. Here, we use the results from the UR group to demonstrate the power-based sample size calculator.  The precision-based demonstration appears in the Supplementary Material. 

To calculate the required sample size, we first estimate the proximal effects of the intervention categories using
only the complete cases in the dataset and we assume \textcolor{red}{$\tau=$}$100\%$ availability as reasoned in Section \ref{s:MLMRT}.
For each participant, we delete the days when the messages were not sent due to technical errors  or when the outcome measures were not collected due to non-response. 
The technical missingness can be dealt with using imputation \cite{Buuren_Groothuis-Oudshoorn_2011} if desired. 

Here, we give an example using the Motivational Message component that has three intervention categories ($M=3$), 
`benefit', `self-efficacy', and `opportunity', 
all proposed at the beginning of the trial. The randomization probability ($\boldsymbol\pi$) for each category (including the control category) was $0.25$.
According to the working model of $Y_{id}$ described in Section \ref{s:analysis}, we consider the constant ($q=p_m=1$) , linear ($q=p_m=2$) and quadratic ($q=p_m=3$) trends for the intervention categories, where $m=1, 2, 3$.  
The regression coefficients ($\boldsymbol\beta$) 
can be estimated by equation (\ref{theta_hat}). We denote the initial and average proximal effect sizes by $\boldsymbol\beta^0$ and $\bar{\boldsymbol\beta}^d$ respectively, where $\boldsymbol\beta^d$ is defined by equation (\ref{Zbeta_p}). 
The corresponding standardized effect sizes are denoted by $\boldsymbol\delta^0$ ($\boldsymbol\beta^0/\bar{\sigma}$) and $\bar{\boldsymbol\delta}^d$ ($\bar{\boldsymbol\beta}^d/\bar{\sigma}$) respectively, where $\bar{\sigma}^2$ is the average of the residual variance over all the decision time points. 

Under the constant trend assumption, the initial and average proximal effect sizes
are the same, because the effect sizes are marginalized over the study days and the historical variables are not considered in the model. 
We have  $\boldsymbol\beta=(357, 589, 526)^{\top}$=$\boldsymbol\beta^0$=$\bar{\boldsymbol\beta}^d$ and $\bar{\sigma}$=$4869$ or $\boldsymbol\delta^0$=$\bar{\boldsymbol\delta}^d$=$(0.073, 0.121, 0.108)^{\top}$.
Assuming $100\%$ availability, $44$ decision time points, type-I error rate $\alpha=5\%$, $80\%$ power, and using 
the more conservative Hotelling's $T^2$ distributed test statistic in equation (\ref{N_F_N_q_1_P}), the sample size required to detect a three-category component with average standardized proximal effect sizes $0.073$, $0.121$ and $0.108$ is $117$. 
This calculation approach was mentioned by Figureroa \textit{et al} \cite{Figueroa_etal_2020b}. 

Assuming linear and quadratic trends, we have the initial proximal effect size vectors $(609, 441, 869)^{\top}$ and $(662, 718, 1394)^{\top}$, the average proximal effect size vectors $(338, 598, 512)^{\top}$ and $(378, 621, 530)^{\top}$, $\bar{\sigma}$=$4867$ and $4866$, respectively. The corresponding standardized initial proximal effect size vectors are $(0.125, 0.091, 0.178)^{\top}$ and $(0.136, 0.148, 0.287)^{\top}$, while the corresponding standardized average proximal effect size vectors are $(0.069, 0.123, 0.105)^{\top}$ and $(0.078, 0.128, 0.109)^{\top}$, respectively. 
We observed the quadratic trend, where the shape of the proximal effect size of the `benefit' category is concave down with local maximum at the 36-th decision time point 
while both the `self efficacy' and `opportunity' categories are concave up with local minimum at the 17-th and 26-th decision time points, respectively.
Therefore, 
the calculated sample sizes corresponding to the linear and quadratic trends are $116$ and $101$ respectively.
The required sample size under the quadratic trend turns out to be the smallest 
because the estimated initial and average proximal effect sizes using the quadratic trend are larger than the effect sizes using the constant and linear trends, given the same length of study periods.

Suppose it is reasonable to assume that the effects of the intervention categories are roughly equal, or that the interest lies in the `pooled' effect of all the intervention categories. We may use the conventional two-category approach \cite{Liao_etal_2016}, where the components are collapsed into e.g., no message vs a `pooled' intervention message. The estimated standardized average proximal effect size for the Motivational Message component will be $0.101$, 
under a constant trend, and the corresponding sample size required will be 72. However, the equal intervention category effects assumption may not hold and the interest may be in identifying the individual effect sizes. Then the multi-category approach as demonstrated earlier is more useful. 

Suppose the smaller study considered adding more intervention categories later in the study. 
For example, suppose two additional categories are added to the Motivational Message component at halfway (i.e., the 23rd decision point), based on the feedback received from the participants since the trial started.
Suppose we use the same proximal effect sizes for the first three categories and $\bar{\sigma}$ as estimated above, and assume the proximal effect sizes of both added categories follow constant trends with average value $300$ and $100\%$ availability, the Hotelling's $T^2$ distributed test statistic calculates a sample size of $163$, for $\alpha = 5\%$ and $80\%$ power. 
Suppose the availability is not always $100\%$; for example, some participants may turn off the activity notification on some particular days. The calculated sample size increases to $230$ and $319$ for $70\%$ and $50\%$ expected availabilities, respectively.

The data analysis and the sample size calculation results described above are summarized in Table \ref{Table: DIAMANTE_student} below.
\begin{table}[h]
\centering
\caption{
The sample sizes calculation based on the data analysis results of the DIAMANTE study with the university students.
}
\label{Table: DIAMANTE_student}
\centering
\begin{tabular}{rllrrrrrrr}
  \hline
& $\tau$ & Trend & Category & $\boldsymbol\beta^0$ & $\bar{\boldsymbol\beta}^d$ & $\bar{\sigma}$ & $\boldsymbol\delta^0$ & $\bar{\boldsymbol\delta}^d$ & $N$ \\ 
  \hline
& 100\% & Constant & Benefit          & 357  & 357 & 4869 & 0.073 & 0.073 & 117 \\ 
&            &                & Self-Efficacy & 589  & 589 &          & 0.121 & 0.121 &  \\ 
&            &                & Opportunity & 526  & 526 &          & 0.108 & 0.108 &  \\ 
&            & Linear     & Benefit           & 609 & 338 & 4867 & 0.125 & 0.069 & 116 \\ 
&            &                & Self-Efficacy & 441 & 598 &          & 0.091 & 0.123 &  \\ 
&            &                & Opportunity & 869 & 512 &          & 0.178 & 0.105 &  \\ 
&            & Quadratic & Benefit           & 662 & 378 & 4866 & 0.136 & 0.078 & 101 \\ 
&            &                 & Self-Efficacy & 718 & 621 &          & 0.148 & 0.128 &  \\ 
&            &                 & Opportunity & 1394 & 530 &        & 0.287 & 0.109 &  \\ 
 \\
&            & Constant  & Pooled        & 494 & 494 & 4870 & 0.101 & 0.101 & 72 \\
 \\
&            & Constant & Benefit          & 357  & 357 & 4869 & 0.073 & 0.073 & 163 \\ 
&            &                & Self-Efficacy & 589  & 589 &          & 0.121 & 0.121 &  \\ 
&            &                & Opportunity & 526  & 526 &          & 0.108 & 0.108 &  \\ 
&            &                & Message 4  & 300 & 300 &          & 0.062 & 0.062 &  \\
&            &                & Message 5  & 300 & 300 &          & 0.062 & 0.062 &  \\
\\
&   70\% & Constant & Benefit          & 357  & 357 & 4869 & 0.073 & 0.073 & 230 \\ 
&            &                & Self-Efficacy & 589  & 589 &          & 0.121 & 0.121 &  \\\
&            &                & Opportunity & 526  & 526 &          & 0.108 & 0.108 &  \\ 
\\
&   50\% & Constant & Benefit          & 357  & 357 & 4869 & 0.073 & 0.073 & 319 \\ 
&            &                & Self-Efficacy & 589  & 589 &          & 0.121 & 0.121 &  \\\
&            &                & Opportunity & 526  & 526 &          & 0.108 & 0.108 &  \\ 
    \hline
\end{tabular}
\end{table}

\section{Simulation Study
}\label{s:sim}
In this section, we present simulation studies to investigate the performance of the proposed power-based sample size ($N$) formulas.
Let the study period be $D=180$ and the number of decision time points per day be $T = 1$. Let there be a control category, and $M = 4$ intervention categories, where $M_0 = 3$ categories are added at the beginning, and $M_1 = 1$ category is added halfway through the study at $d_1 = 91$. 
We calculate $N$ under the correctly specified or some mis-specific models for each simulation study, with nominal power $P=80\%$ and Type-I error rate $\alpha = 5\%$, and generate $1000$ Monte Carlo (MC) data sets. 
The performance of each $N$ formula is measured by comparing the difference between the formulated and MC Power estimates.
The data generation steps are listed below.
\begin{enumerate}
\item[Step 1.] The availability indicators $I_{id}$ follow the Bernoulli distributions, i.e. $I_{id}\sim \text{Bernoulli}(\tau_{d})$ for each participant $i$ and on each day $d$. 
We set the availability rate $\tau_{d}$ at 100\% and 70\%.
\item[Step 2.] The intervention categories $\boldsymbol A_{id}$=$\left (A_{i1d}, \ldots, A_{i( \sum_{j=0}^{k}M_j )d}\right )^{\top}$ 
follow the
multinomial 
distributions, i.e. $\boldsymbol A_{id}\sim$Multinomial$\left (1-\sum_m\pi_{md}, \pi_{1d}, \ldots, \pi_{( \sum_{j=0}^{k}M_j  )d}\right )$, 
where $\pi_{md}$ and $1-\Sigma_m \pi_{md}$ are the randomization probabilities of the intervention category-$m$ and the control category on day $d$, respectively. 
We set the initial randomization probability as  $\pi_{md} = 0.25$, and $\pi_{md} = 0.2$ after $d_1$. 
\item[Step 3.] The error terms $\boldsymbol\epsilon_{i}$ = ($\epsilon_{i1}$,$\ldots$,$\epsilon_{iD}$)$^{\top}$ 
follow
the multivariate normal distributions, i.e., $\boldsymbol\epsilon_{i}\sim$MVN($\boldsymbol 0_{D\times 1}$, COV($\boldsymbol\epsilon_{i}$)), where COV($\boldsymbol\epsilon_{i}$) is the $D\times D$ dimensional covariance matrix with diagonal entries $\sigma^2$ and off-diagonal entries $\rho\sigma^2$.
We set $\sigma$=1 and $\rho$=0.
\item[Step 4.] The proximal outcome is computed by $Y_{id}$ = $\boldsymbol X_{id}^{\top}\boldsymbol\theta$ + $\epsilon_{id}$, where $\boldsymbol\theta$=($\boldsymbol\alpha^\top$, $\boldsymbol\beta^\top$)$^\top$
and $\boldsymbol X_{id}^{\top}$=[$\boldsymbol B_{d}^{\top}$, 
$( A_{i1d} - \pi_{1d} )\boldsymbol Z_{1d}^{\top}$, $\ldots$, $( A_{i4d} - \pi_{4d} )\boldsymbol Z_{4d}^{\top}$].
We set $\boldsymbol B_{d} = \left(1, \text{min}\left[28-1, d - 1 \right] \right)$, and consider a linear-plateau trend for the standardized proximal effect size for each intervention category with turning point on the 28-th day, i.e.,
$d^{m}_{\text{turn}}$=28 for
$m$=1, 2 and 3 and $d^{m}_{\text{turn}}=118$ for $m$=4.
\end{enumerate}
The standardized proximal effect size of the m-th category satisfies $\delta_m(d, \boldsymbol\delta_m)=\boldsymbol Z_{md}^{\top}\boldsymbol \delta_m$, where $\boldsymbol \delta_m=\boldsymbol\beta_m/\bar{\sigma}$.
We define
the initial standardized proximal effect sizes 
to be $0.001$ and
the average standardized proximal effect sizes 
to be $0.1$ and $0.06$.

\subsection{Correctly specified working model}\label{s:correct model}
Table \ref{Table: T1} gives the calculated $N$, and the corresponding formulated and MC Power estimates.
We observe that the Power estimates are fairly similar.  
In general, when the working model is correctly specified, the proposed sample size calculator provides accurate Power. 

\begin{table}[h]
\centering
\caption{
The $N$ and Power estimates
when the working model assumptions are correct. 
}
\label{Table: T1}
\begin{tabular}{rllrrrrrr}
  \hline
  & \multicolumn{2}{c}{} & \multicolumn{6}{c}{Average standardized proximal effect size} \\
 & \multicolumn{2}{c}{} & 0.10 & 0.06 & 0.10 & 0.06 & 0.10 & 0.06 \\
 & Availability & Test Statistics  & \multicolumn{2}{c}{$N$} & \multicolumn{2}{c}{Formulated Power}  & \multicolumn{2}{c}{Monte Carlo Power}  \\ 
  \hline
 & 100\% & $\chi^2_{\sum_{m}p_m}$                          &  46 & 127 & 0.81 & 0.80 & 0.81 & 0.81 \\ 
 &            & Hotelling's $T^2_{\sum_{m}p_m, N}$        &  54 & 135 & 0.81 & 0.80 & 0.81 & 0.81 \\ 
 &            & Hotelling's $T^2_{\sum_{m}p_m, N-q-1}$  &  54 & 135 & 0.80 & 0.80 & 0.77 & 0.78 \\ 
 & 70\%   & $\chi^2_{\sum_{m}p_m}$                           &  65 & 182 & 0.80 & 0.80 & 0.81 & 0.78 \\ 
 &            & Hotelling's $T^2_{\sum_{m}p_m, N}$        &  73 & 190 & 0.80 & 0.80 & 0.78 & 0.80 \\ 
 &            & Hotelling's $T^2_{\sum_{m}p_m, N-q-1}$   &  73 & 190 & 0.80 & 0.80 & 0.80 & 0.81 \\ 
   \hline
\end{tabular}
\end{table}

\subsection{Mis-specified trend for proximal effect}\label{s:mis trend}
\begin{figure}
\begin{center}
\begin{tabular}{cc}
  \includegraphics[scale=0.3]{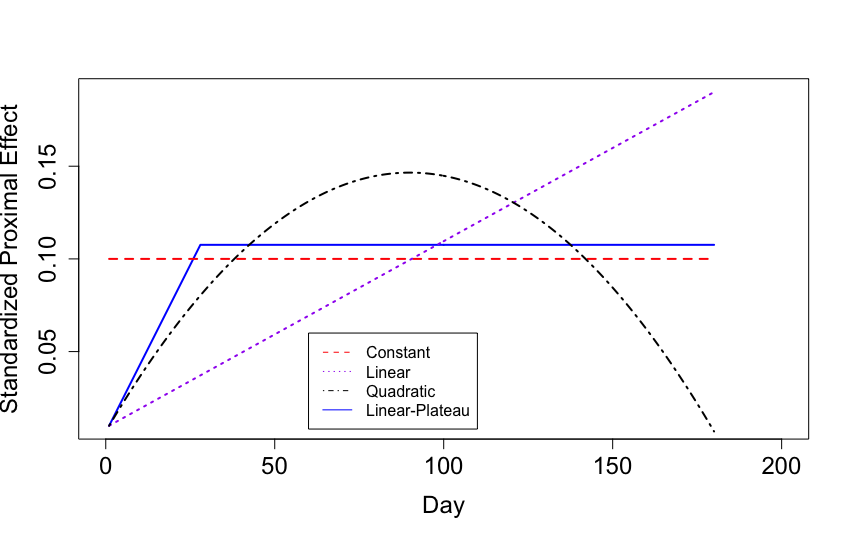}  &   \includegraphics[scale=0.3]{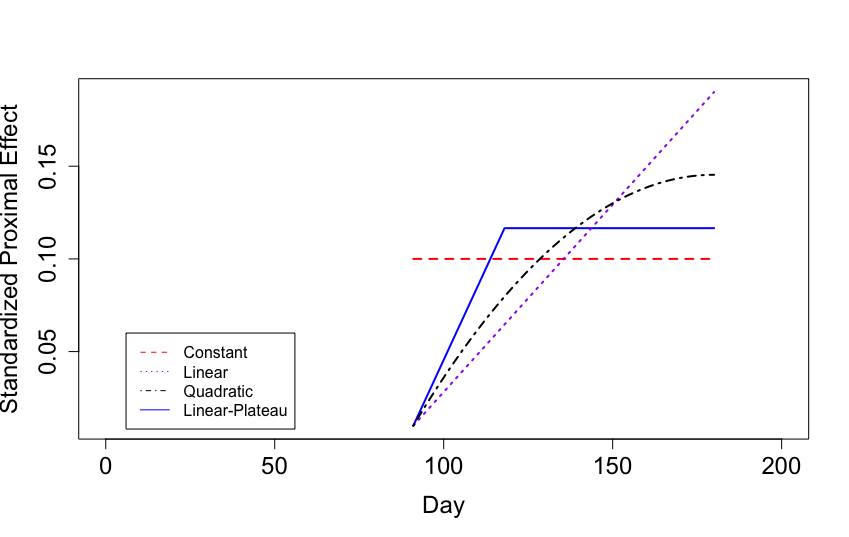} \\
 a) Intervention category proposed at the beginning  & b) Intervention category proposed on the half-way \\
\end{tabular}
\end{center}
\caption{
The plots of standardized proximial effect size over the course of the study period.
}
\label{fig: MisTrend}
\end{figure}

Next, we consider the robustness of the sample size calculator when the model is mis-specified. 
We observe that the formulated Power estimates are very close to the nominal Power, but the MC Power estimates are lower. 
This is because the mis-specified proximal trends give smaller $N$s in contrast to the $N$s in Table \ref{Table: T1}. 
Comparing Tables \ref{Table: T1} and \ref{Table: P_MisTrend}, mis-specifying as the constant trend results in the largest difference in $N$, mainly due to the shape differences (see Figure \ref{fig: MisTrend}). In contrast, the linear and quadratic trends give similar $N$s. 

\begin{table}[h]
\centering
\caption{
The $N$ and Power estimates
when the standardized proximal effect size trends are mis-specified.
}
\label{Table: P_MisTrend}
\begin{tabular}{rlllrrrrrr}
  \hline
 & \multicolumn{3}{c}{} &  \multicolumn{6}{c}{Average standardized proximal effect size}  \\
 & \multicolumn{3}{c}{} & 0.1 & 0.06 & 0.1 & 0.06 & 0.1 & 0.06   \\
 & Trend & Availability & Test Statistics & \multicolumn{2}{c}{$N$} & \multicolumn{2}{c}{Formulated Power}  & \multicolumn{2}{c}{Monte Carlo Power} \\ 
  \hline
& Constant & 100\% & $\chi^2_{\sum_m p_m}$                          &  39 & 107 & 0.81 & 0.80 & 0.73 & 0.69 \\ 
 &               &            & Hotelling's $T^2_{\sum_m p_m, N} $       &  43 & 111 & 0.80 & 0.80 & 0.64 & 0.70 \\ 
 &               &            & Hotelling's $T^2_{\sum_m p_m, N-q-1}$ &  44 & 111 & 0.81 & 0.80 & 0.66 & 0.66 \\ 
 &               & 70\%   & $\chi^2_{\sum_m p_m}$                           &  55 & 152 & 0.80 & 0.80 & 0.70 & 0.68 \\ 
 &               &            & Hotelling's $T^2_{\sum_m p_m, N} $        &  60 & 157 & 0.81 & 0.80 & 0.69 & 0.71 \\ 
 &               &            & Hotelling's $T^2_{\sum_m p_m, N-q-1}$  &  60 & 157 & 0.80 & 0.80 & 0.64 & 0.69 \\ 
\\
& Linear     & 100\% & $\chi^2_{\sum_m p_m}$                                   &  41 & 116 & 0.81 & 0.80 & 0.76 & 0.74 \\ 
 &               &            & Hotelling's $T^2_{\sum_m p_m, N} $                &  49 & 124 & 0.81 & 0.80 & 0.77 & 0.77 \\ 
 &               &            & Hotelling's $T^2_{\sum_m p_m, N-q-1}$   &  49 & 124 & 0.80 & 0.80 & 0.73 & 0.76 \\ 
 &               & 70\%   & $\chi^2_{\sum_m p_m}$ &  58 & 166        & 0.80 & 0.80 & 0.73 & 0.77 \\ 
 &               &             & Hotelling's $T^2_{\sum_m p_m, N} $        &  66 & 174 & 0.80 & 0.80 & 0.74 & 0.75 \\ 
 &               &             & Hotelling's $T^2_{\sum_m p_m, N-q-1}$   &  66 & 174 & 0.80 & 0.80 & 0.73 & 0.73 \\ 
\\
& Quadratic & 100\% & $\chi^2_{\sum_m p_m}$                           &  40 & 115 & 0.80 & 0.80 & 0.74 & 0.73 \\ 
 &                &            & Hotelling's $T^2_{\sum_m p_m, N} $        &  51 & 126 & 0.80 & 0.80 & 0.73 & 0.76 \\ 
 &                &             & Hotelling's $T^2_{\sum_m p_m, N-q-1}$ &  52 & 126 & 0.81 & 0.80 & 0.76 & 0.71 \\ 
 &                &    70\% & $\chi^2_{\sum_m p_m}$                           &  57 & 165 & 0.80 & 0.80 & 0.74 & 0.78 \\ 
 &                &            & Hotelling's $T^2_{\sum_m p_m, N} $        &  68 & 175 & 0.80 & 0.80 & 0.74 & 0.76 \\ 
 &                &            & Hotelling's $T^2_{\sum_m p_m, N-q-1}$ &  69 & 175 & 0.81 & 0.80 & 0.77 & 0.74 \\ 
   \hline
\end{tabular}
\end{table}

\newpage

\subsection{Mis-specified number of intervention categories}\label{s:mis cat}
Suppose we mis-specified $M_1$ such that no categories are added later in the study ($M_1 = 0$) and underestimated the number of initial categories $M_0=1$ in Table \ref{Table: P_MisCat}. The formulated Power estimates are very close to the nominal power, but because the $N$s are calculated assuming fewer categories, the MC Power estimates are much lower than $80\%$. The MC Power estimates (i.e., around 30\% to 40\%) performed the worst when $M_0=1$ and $M_1=0$, as they are the furthest away from the true values (i.e. $M_0=3$ and $M_1=1$). Note that when $M_1 = 0$, the \textit{design structure} essentially reduces to a conventional MRT. However the sample size calculation under a conventional MRT approach will always assume $M_0=1$ by `pooling' the intervention categories into one proximal effect (Liao et al \cite{Liao_etal_2016}), i.e., the sample sizes for $M_0 = 3$ and $M_1= 0$ is the same as that for $M_0=1$ and $M_1=0$.  
The MC Power estimates under $M_0=3$ and $M_1=0$ are around 70\% while the MC Power estimates under $M_0=4$ and $M_1=0$ are around 85\%.
However, in order to detect the proximal effect sizes of the multiple intervention categories that are not all added initially, the proposed method provides a better solution of estimating powers than the Liao's method \cite{Liao_etal_2016}.

\begin{table}[h]
\centering
\caption{
The $N$ and Power estimates
when 
$M_0$ and $M_1$, and consequently $M$ are mis-specified.}
\label{Table: P_MisCat}
\begin{tabular}{rlllrrrrrr}
  \hline
 & \multicolumn{3}{c}{} &  \multicolumn{6}{c}{Average standardized proximal effect size}  \\
 & \multicolumn{3}{c}{} & 0.1 & 0.06 & 0.1 & 0.06 & 0.1 & 0.06 \\
 & $M_0$ and $M_1$ & Availability & Test Statistics &  \multicolumn{2}{c}{$N$} & \multicolumn{2}{c}{Formulated Power}  & \multicolumn{2}{c}{Monte Carlo Power} \\ 
  \hline
  & $M_0=1$ and $M_1=0$ & 100\% & $\chi^2_{\sum_m p_m}$                           &  21 & 58 & 0.81 & 0.81 & 0.42 & 0.39 \\ 
  &                                        &            & Hotelling's $T^2_{\sum_m p_m, N}$        &  24 & 61 & 0.81 & 0.81 & 0.30 & 0.34 \\ 
  &                                        &            & Hotelling's $T^2_{\sum_m p_m, N-q-1}$  &  24 & 61 & 0.80 & 0.80 & 0.26 & 0.36 \\ 
  &                                        & 70\%   & $\chi^2_{\sum_mp_m}$                            &  30 & 82 & 0.81 & 0.80 & 0.41 & 0.41 \\ 
  &                                        &            & Hotelling's $T^2_{\sum_m p_m, N}$         &  33 & 85 & 0.81 & 0.80 & 0.31 & 0.38 \\ 
   &                                       &            & Hotelling's $T^2_{\sum_m p_m, N-q-1}$  &  33 & 85 & 0.80 & 0.80 & 0.29 & 0.38 \\ 
  \\
  & $M_0=3$ and $M_1=0$ & 100\% & $\chi^2_{\sum_m p_m}$                          &  39 & 109 & 0.80 & 0.80 & 0.71 & 0.73 \\ 
  &                                        &            & Hotelling's $T^2_{\sum_m p_m, N}$       &  46 & 115 & 0.81 & 0.80 & 0.68 & 0.71 \\ 
  &                                        &            & Hotelling's $T^2_{\sum_m p_m, N-q-1}$  &  46 & 115 & 0.81 & 0.80 & 0.68 & 0.70 \\ 
  &                                        & 70\%   & $\chi^2_{\sum_m p_m}$                           &  56 & 155 & 0.81 & 0.80 & 0.73 & 0.75 \\ 
  &                                        &            & Hotelling's $T^2_{\sum_m p_m, N}$        &  62 & 161 & 0.80 & 0.80 & 0.71 & 0.71 \\ 
  &                                        &            & Hotelling's $T^2_{\sum_m p_m, N-q-1}$  &  62 & 161 & 0.80 & 0.80 & 0.71 & 0.72 \\ 
  \\
  & $M_0=4$ and $M_1=0$ &  100\% & $\chi^2_{\sum_m p_m}$                          &  50 & 140 &  0.80 &  0.80 & 0.84 & 0.85 \\ 
  &                                        &            &  Hotelling's $T^2_{\sum_m p_m, N}$       &   58 & 148 &  0.80 &  0.80 & 0.84 & 0.85 \\ 
  &                                        &            &  Hotelling's $T^2_{\sum_m p_m, N-q-1}$ &  59 & 148 &  0.81 & 0.80 & 0.82 & 0.86 \\ 
  &                                        &  70\%   &  $\chi^2_{\sum_m p_m}$                          & 72 &  200 & 0.80 & 0.80 & 0.86 &  0.84 \\ 
  &                                        &            &  Hotelling's $T^2_{\sum_m p_m, N}$        &  80 &  208 & 0.81 & 0.80 & 0.84 &  0.81 \\ 
  &                                        &            &  Hotelling's $T^2_{\sum_m p_m, N-q-1}$ &  80 &  208 &  0.80 &  0.80 & 0.84 & 0.85 \\ 
   \hline
\end{tabular}
\end{table}

\subsection{Mis-specified error term distribution}
We calculate the $N$s and the corresponding Powers
assuming that the error terms 
are generated from the multivariate normal distribution, 
i.e., $\boldsymbol\epsilon_{i}\sim$MVN($\boldsymbol 0_{D\times 1}$, COV($\boldsymbol\epsilon_{i}$)), 
where 
a constant variance $\sigma^2$ is used in the $D\times D$ dimensional covariance matrix. However, for the data generation, we let the true variance of $\epsilon_{id}$ to be $\sigma^2_{d}$, for $d$=$1,\ldots,D$, which varies over $d$. 
For example, $\sigma^2_d$ is linearly increasing (i.e. $\sigma^2_{d}=0.9^2+0.0021\times (d-1)$) and decreasing (i.e. $\sigma^2_{d}=1.19^2-0.0021\times (d-1)$), where $\bar{\sigma}^2=\dfrac{1}{D}\sum_{d=1}^D\sigma^2_{d}=1$. 
We observe that both the formulated and MC Power estimates remain close to the nominal Power in Table \ref{Table: P_MisSigma}, meaning our
sample size calculator is still accurate when the outcome variances are not constant over the study.  

\begin{table}[h]
\centering
\caption{
The $N$ and Power estimates
when $\sigma_d^2$ is not constant over $d$.
}
\label{Table: P_MisSigma}
\begin{tabular}{rlllrrrrrr}
  \hline
 & \multicolumn{3}{c}{} &  \multicolumn{6}{c}{Average standardized proximal effect size} \\
 & \multicolumn{3}{c}{} & 0.1 & 0.06 & 0.1 & 0.06 & 0.1 & 0.06  \\
 & Trend of $\sigma_d^2$ & Availability & Test Statistics &   \multicolumn{2}{c}{$N$} & \multicolumn{2}{c}{Formulated Power}  & \multicolumn{2}{c}{Monte Carlo Power} \\ 
  \hline
& Increasing & 100\% & $\chi^2_{\sum_m p_m}$                          &  46 & 127 & 0.81 & 0.80 & 0.80 & 0.80 \\ 
&                   &            & Hotelling's $T^2_{\sum_m p_m, N}$       &  54 & 135 & 0.81 & 0.80 & 0.78 & 0.77 \\ 
&                   &            & Hotelling's $T^2_{\sum_m p_m, N-q-1}$ &  54 & 135 & 0.80 & 0.80 & 0.77 & 0.80 \\ 
&                   & 70\%   & $\chi^2_{\sum_m p_m}$                          &  65 & 182 & 0.80 & 0.80 & 0.78 & 0.80 \\ 
&                   &            & Hotelling's $T^2_{\sum_m p_m, N}$        &  73 & 190 & 0.80 & 0.80 & 0.78 & 0.78 \\ 
&                   &            & Hotelling's $T^2_{\sum_m p_m, N-q-1}$  &  73 & 190 & 0.80 & 0.80 & 0.78 & 0.78 \\ 
  \\
 & Decreasing & 100\% & $\chi^2_{\sum_m p_m}$                         &  46 & 127 & 0.81 & 0.80 & 0.81 & 0.80 \\ 
 &                    &            & Hotelling's $T^2_{\sum_m p_m, N}$       &  54 & 135 & 0.81 & 0.80 & 0.78 & 0.79 \\ 
 &                    &            & Hotelling's $T^2_{\sum_m p_m, N-q-1}$ &  54 & 135 & 0.80 & 0.80 & 0.79 & 0.80 \\ 
 &                    & 70\%   & $\chi^2_{\sum_m p_m}$                         &  65 & 182 & 0.80 & 0.80 & 0.82 & 0.80 \\ 
 &                    &            & Hotelling's $T^2_{\sum_m p_m, N}$       &  73 & 190 & 0.80 & 0.80 & 0.78 & 0.81 \\ 
 &                    &            & Hotelling's $T^2_{\sum_m p_m, N-q-1}$ &  73 & 190 & 0.80 & 0.80 & 0.78 & 0.81 \\    
 \hline
\end{tabular}
\end{table}

Suppose the error term follows a multivariate normal distribution with standardized normal for marginal distribution and non-zero correlation coefficients, i.e., $\boldsymbol\epsilon_{i}$= ($\epsilon_{i1}$,$\ldots$,$\epsilon_{iD}$)$^{\top}$ follows a normal AR(1) process, e.g., $\epsilon_{id}=\phi\epsilon_{i, d-1} + \nu_{id}$, where $\phi$=$0.5$ and $-0.5$ and $\nu_{d}$ are i.i.d Normal(0, 0.75).
When calculating $N$ assuming no autocorrelation, we observe that both the formulated and MC Power estimates are very close to the nominal power in Table \ref{Table: P_MisAR}. 
Therefore, the required $N$ seems robust against the autocorrelation structure of the outcome measures over time.  

\begin{table}[h]
\sf\centering
\caption{
The $N$ and Power estimates
when $\epsilon_{id}=\phi\epsilon_{i, d-1} + \nu_{id}$, where $\phi$=$0.5$ and $-0.5$ and $\nu_{d}$ are i.i.d Normal(0, 0.75).
}
\label{Table: P_MisAR}
\begin{tabular}{rlllrrrrrr}
  \hline
 & \multicolumn{3}{c}{} &  \multicolumn{6}{c}{Average standardized proximal effect size} \\
 & \multicolumn{3}{c}{} & 0.1 & 0.06 & 0.1 & 0.06 & 0.1 & 0.06 \\
 & $\phi$ & Availability & Test Statistics &  \multicolumn{2}{c}{$N$} & \multicolumn{2}{c}{Formulated Power}  & \multicolumn{2}{c}{Monte Carlo Power} \\ 
  \hline
 & $0.5$ & 100\% & $\chi^2_{\sum_m p_m}$                          & 46 & 127 & 0.81 & 0.80 & 0.82 & 0.82 \\ 
 &           &            & Hotelling's $T^2_{\sum_m p_m, N}$        & 54 & 135 & 0.81 & 0.80 & 0.78 & 0.77 \\ 
 &           &            & Hotelling's $T^2_{\sum_m p_m, N-q-1}$ &  54 & 135 & 0.80 & 0.80 & 0.78 & 0.80 \\ 
 &           & 70\%   & $\chi^2_{\sum_m p_m}$                          &  65 & 182 & 0.80 & 0.80 & 0.79 & 0.81 \\ 
 &           &            & Hotelling's $T^2_{\sum_m p_m, N}$        &  73 & 190 & 0.80 & 0.80 & 0.79 & 0.78 \\ 
 &           &            & Hotelling's $T^2_{\sum_m p_m, N-q-1}$  &  73 & 190 & 0.80 & 0.80 & 0.77 & 0.79 \\ 
  \\
& $-0.5$ & 100\% & $\chi^2_{\sum_m p_m}$                           & 46 & 127 & 0.81 & 0.80 & 0.80 & 0.81 \\ 
&            &             & Hotelling's $T^2_{\sum_m p_m, N}$       &  54 & 135 & 0.81 & 0.80 & 0.83 & 0.79 \\ 
&            &             & Hotelling's $T^2_{\sum_m p_m, N-q-1}$ &  54 & 135 & 0.80 & 0.80 & 0.80 & 0.77 \\ 
 &           & 70\%    & $\chi^2_{\sum_m p_m}$                          &  65 & 182     & 0.80 & 0.80 & 0.80 & 0.82 \\ 
 &           &             & Hotelling's $T^2_{\sum_m p_m, N}$       &  73 & 190 & 0.80 & 0.80 & 0.78 & 0.82 \\ 
 &           &             & Hotelling's $T^2_{\sum_m p_m, N-q-1}$ &  73 & 190 & 0.80 & 0.80 & 0.79 & 0.78 \\ 
   \hline
\end{tabular}
\end{table}

\subsection{Mis-specified trend of availability}\label{s:mistau}
It is common to calculate $N$ under a simple assumption of constant trend for availability. However, the availability trend may not actually be constant in real data. Here, we calculate $N$ assuming a constant trend of $70\%$ availability ($\tau_d = 0.7$). 
The datasets are however generated based on linearly increasing (i.e. $\tau_{d}=0.5+0.0022\times (d-1)$ ) and decreasing (i.e. $\tau_{d}=0.9-0.0022\times (d-1)$) availability trends, for $d$=$1,\ldots,D$, where we have $\bar{\tau}=\dfrac{1}{D}\sum_{d=1}^D\tau_{d}=0.7$. 
In Table \ref{Table: P_MisTau}, both the formulated and MC Power estimates are close to the nominal Power, showing that mis-specifying the availability rate trend over time has little impact on the Power, and consequently the required $N$. 

\begin{table}[h]
\centering
\caption{
The $N$ and Power estimates
when $\tau_d$ is not constant over $d$, i.e. linearly increasing or decreasing over $d$.
}
\label{Table: P_MisTau}
\begin{tabular}{rllrrrrrr}
  \hline
 & \multicolumn{2}{c}{} &  \multicolumn{6}{c}{Average standardized proximal effect size}  \\
 & \multicolumn{2}{c}{} & 0.1 & 0.06 & 0.1 & 0.06 & 0.1 & 0.06 \\
 & Trend of $\tau_d$ & Test Statistics &   \multicolumn{2}{c}{$N$} & \multicolumn{2}{c}{Formulated Power}  & \multicolumn{2}{c}{Monte Carlo Power} \\ 
  \hline
& Increasing  & $\chi^2_{\sum_m p_m}$                           &  65 & 182 & 0.80 & 0.80 & 0.80 & 0.79 \\ 
&                   & Hotelling's $T^2_{\sum_m p_m, N}$         &  73 & 190 & 0.80 & 0.80 & 0.82 & 0.79 \\ 
&                   & Hotelling's $T^2_{\sum_m p_m, N-q-1}$   &  73 & 190 & 0.80 & 0.80 & 0.80 & 0.78 \\
  \\
 & Decreasing & $\chi^2_{\sum_m p_m}$                          &  65 & 182 & 0.80 & 0.80 & 0.79 & 0.79 \\ 
 &                    & Hotelling's $T^2_{\sum_m p_m, N}$        &  73 & 190 & 0.80 & 0.80 & 0.78 & 0.77 \\ 
 &                    & Hotelling's $T^2_{\sum_m p_m, N-q-1}$  &  73 & 190 & 0.80 & 0.80 & 0.77 & 0.80 \\  
 \hline
\end{tabular}
\end{table}

\section{Discussion}\label{s:disc}
In this paper, we propose a novel FlexiMRT design wherein it is possible to allow categories of intervention components to be added not only at the beginning, but also later in the study, with the adding times and the numbers of categories to be added at these times being pre-determined. 
We derive the associated sample size calculation methods based on both power and precision of the proximal effect size estimates.
For both the methods, the required sample sizes increase with the number of intervention categories, 
and decrease when either the study period is longer or the number of decision time points per day is larger. 
We also observe that the required sample size decreases when more categories are added later in the study than at the beginning, given a fixed total number of intervention categories and assuming constant proximal effect trends. 
The proposed methods give the Monte Carlo estimates of power and coverage probability close to the corresponding nominal values, provided that the specified working model is ``not too far" from the true one. 
A sample size calculated by incorrectly using the conventional MRT sample size calculator that only considers two-category components \cite{Liao_etal_2016} does not provide sufficient power to detect the proximal effects. 

As mentioned, the DIAMANTE study is the primary motivation behind this paper. 
We have shown how the proposed sample size calculation method can be applied for the UR group of the DIAMANTE student study. However, it is worth noting that the calculated sample size is aimed not to select the optimal category, but to detect whether at least one of the intervention categories is more effective than the control category.

We are currently collecting data through the main DIAMANTE study, and performing interim data analysis, which can be used to further improve the experimental design. 
When the data collection is complete, we can apply linear mixed model approach to investigate the between-person heterogeneity of the proximal effect of an intervention component. This is similar to the method of Qian \textit{et al} \cite{qian_etal_2020}.  
Based on the results of Bidargaddi \textit{et al} \cite{Bidargaddi_etal_2018}, we could investigate, for example, whether a message category is more effective at mid-day on weekends than other decision time points. 

The proposed FlexiMRT design only allows a pre-specified number of new intervention categories to be added at specified time points during the study.
However, no additional participants are recruited when new categories are added.
These features distinguish FlexiMRT from the platform clinical trials \cite{Ventz_etal_2017},
where the number of treatment arms added and the corresponding times are not known a priori, and additional participants are recruited when more arms are added.
The proposed method, therefore, requires adequate planning to pre-determine the number of intervention categories and when to add them. 
Extending the procedure to be able to add arbitrary new categories into the FlexiMRT design can be an important future direction.
This design can be further extended using a decision-theoretic framework similar to Lee \textit{et al} \cite{Lee_etal_2019} to investigate when to add or not add a message category based on the observed proximal outcomes. 
Alternatively, a message category can be dropped early if it is unlikely to have any effect.
The study can be stopped if the efficacy of a category is recognized early. 
These approaches suggested by Magirr \textit{et al} \cite{Magirr_etal_2012} can be used to evaluate the intervention categories of an MRT design efficiently through a series of interim analyses.

The proposed sample size calculators can be extended to allow for adaptive randomization probabilities, determined by the proximal outcomes and message categories at previous decision time points \cite{Wason_Trippa_2014}, with the aim of sending more effective messages to the participants. 
In other words, the sequential outcomes and randomization probabilities of the current message categories would depend on the outcomes and message categories from previous decision time points, in order to achieve experimental objectives. 
In this type of design, we can estimate not only the current proximal effect, but also the delayed effect, of each category of a particular intervention component. 
Similar to Dempsey \textit{et al} \cite{Dempsey_etal_2020}, one can also consider stratifying strategies, where the inverse probability weighting techniques may be incorporated into the test statistic construction, enabling it to deal with more complex dependencies, e.g., a setting where the proximal outcome may depend on both treatments of today and yesterday.
Another possible future direction can be extending the sample size calculators to also account for the binary outcomes, similar to the method of Qian \textit{et al} \cite{Qian_etal_2021}.

\section*{Data Availability Statement}
The students dataset from the adapted DIAMANTE study used to demonstrate the proposed sample size calculators in Section \ref{s:realexample} is available from the co-author, Dr. Caroline Figueroa, \texttt{C.Figueroa@tudelft.nl}, upon reasonable request.

\section*{Declaration of Conflicting Interest}
The authors declare no potential conflict of interests.

\section*{Acknowledgement}
The authors would like to thank the reviewers and editors for the valuable comments and feedback. We also acknowledge the feedback from our colleague, Dr. Raju Maiti. 
We would like to acknowledge Chris Karr who helped developing and managing the DIAMANTE automated text messaging service and passive data collection.
This work has been partially supported by Khoo Bridge Funding Award (Duke-NUS-KBrFA/2021/0040) and the start-up grant from the Duke-NUS Medical School, Singapore, as well as an Academic Research Fund Tier 2 grant (MOE-T2EP20122-0013) from the Ministry of Education to Dr. Bibhas Chakraborty. 
The DIAMANTE trial has been funded by an R01 grant to Dr. Adrian Aguilera (University of California, Berkeley) and Dr. Courtney Lyles (University of California, San Francisco) who designed the DIAMANTE application, 1R01 HS25429-01 from the Agency for Healthcare Research and Quality. 
We acknowledge Dr Adrian Aguilera and Dr Courtney Lyles, the Principal Investigators of the DIAMANTE study, for involving us in this important mHealth study and for sharing the UC-Berkeley student data.

\bibliography{wileyNJD-AMA}%

\appendix

\section{Derivation of the covariance matrix of $\hat{\boldsymbol\beta}$}\label{app: Sigma_beta}
This section derives the covariance matrix of the parameter estimator $\hat{\boldsymbol\beta}$. 
First, we derive the expression for equation (\ref{theta_tilde_aa}), i.e. $ \tilde{\boldsymbol\theta}$=$\left[\sum_{d=1}^{D}E\left( I_{id} \boldsymbol X_{id} \boldsymbol X_{id}^{\top} \right)\right]^{-1} \sum_{d=1}^{D}E\left( I_{id}Y_{i, d}\boldsymbol X_{id} \right)$, in more detail.
The expectation within the second summation can be expressed by 
\begin{align*}
&E\left( I_{id} Y_{i, d} \boldsymbol X_{id} \right)\\
=&\begin{bmatrix} 
&E( I_{ i d } Y_{ i, d } \boldsymbol B_{d} ) \\
&E( I_{ i d } Y_{ i, d }( A_{i1d}-\pi_{1d} )\boldsymbol Z_{1d} )\\
&\vdots\\
&E( I_{ i d }Y_{ i, d }( A_{ i M_{ 0 } d } - \pi_{ M_{ 0 } d } ) \boldsymbol Z_{M_{ 0 }d} )\\
&\vdots\\
&E( I_{ i d }Y_{ i, d }( A_{ i ( \sum_{j=0}^{k-1} M_{ j } + 1 ) d } - \pi_{ ( \sum_{j=0}^{k-1} M_{ j } + 1 ) d } ) \boldsymbol Z_{( \sum_{j=0}^{k-1} M_{ j } + 1 )d} ) \\
&\vdots\\
&E( I_{ i d }Y_{ i, d }( A_{ i ( \sum_{j=0}^{k} M_{ j }  ) d } - \pi_{ ( \sum_{j=0}^{k} M_{ j }  ) d } ) \boldsymbol Z_{ ( \sum_{j=0}^{k} M_{ j }  )d} ) \\
\end{bmatrix}\\
=&\begin{bmatrix}
& E( I_{ i d })E( Y_{ i, d } \boldsymbol B_{d} ) \\
& E( I_{ i d })E( Y_{ i, d }(A_{i1d}-\pi_{1d}) )\boldsymbol Z_{1d} \\
& \vdots \\
& E( I_{ i d })E( Y_{ i, d }( A_{ i ( \sum_{j=0}^{k} M_{ j }  ) d } - \pi_{ ( \sum_{j=0}^{k} M_{ j }  ) d } ) ) \boldsymbol Z_{ ( \sum_{j=0}^{k} M_{ j } ) d} \\
\end{bmatrix}\\
=&\begin{bmatrix} 
& \tau_{d} \boldsymbol B_{d}^{\top}\boldsymbol\alpha\boldsymbol B_{d} \\
& \tau_{d} E( Y_{ i, d }(A_{i1d}-\pi_{1d}) )\boldsymbol Z_{1d} \\
& \vdots \\
& \tau_{d} E( Y_{ i, d }( A_{ i ( \sum_{j=0}^{k} M_{ j }  ) d } - \pi_{ ( \sum_{j=0}^{k} M_{ j }  ) d } ) )\boldsymbol Z_{( \sum_{j=0}^{k} M_{ j }  ) d} \\
\end{bmatrix}
\end{align*}
\normalsize
where 
\begin{align*}
&E( Y_{i, d} ( A_{i1d} - \pi_{1d} ) ) \\
=& E\{ \boldsymbol B_{d}^{\top}\boldsymbol\alpha ( A_{i1d} - \pi_{1d} ) \\
+& \left[ (A_{i1d}-\pi_{1d})\boldsymbol Z_{1d}^{\top}\boldsymbol\beta_1+ \cdots +(A_{iM_{0}d}-\pi_{M_{0}d})\boldsymbol Z_{M_{0}d}^{\top}\boldsymbol\beta_{M_0} \right] ( A_{i1d} - \pi_{1d} ) \\
+& \left[ (A_{i( M_0 + 1 )d}-\pi_{( M_0 + 1 )d})\boldsymbol Z_{( M_0 + 1 )d}^{\top}\boldsymbol\beta_{ ( M_0 + 1 ) }+ \cdots +(A_{i( M_{0} + M_{1} ) d}-\pi_{( M_{0} + M_{1} )d})\boldsymbol Z_{( M_0 + M_{1} )d}^{\top}\boldsymbol\beta_{( M_{0} + M_{1} )} \right]( A_{i1d} - \pi_{1d} ) \\
&\vdots\\
+&\left[ (A_{i( \sum_{j=0}^{k-1} M_j +1) d}-\pi_{( \sum_{j=0}^{k-1}M_j + 1 )d})\boldsymbol Z_{( \sum_{j=0}^{k-1}M_j + 1 )d}^{\top}\boldsymbol\beta_{ ( \sum_{j=0}^{k-1}M_j + 1 ) }+ \cdots +(A_{i( \sum_{j=0}^{k}M_j ) d}-\pi_{( \sum_{j=0}^{k}M_j )d})\boldsymbol Z_{( \sum_{j=0}^{k}M_j )d}^{\top}\boldsymbol\beta_{( \sum_{j=0}^{k}M_j )} \right]\\ 
&( A_{i1d} - \pi_{1d} ) \\
+&\epsilon_{id} ( A_{i1d} - \pi_{1d} ) \} \\
=& \left[ \pi_{ 1 d }(1 - \pi_{ 1 d } )\boldsymbol Z_{1d}^{\top}\boldsymbol\beta_1- \cdots - \pi_{ 1 d }\pi_{ M_{0} d }\boldsymbol Z_{M_{0}d}^{\top}\boldsymbol\beta_{M_0} \right]\\
+& \left[  - \pi_{ 1 d }\pi_{ ( M_0 + 1 ) d } \boldsymbol Z_{( M_0 + 1 )d}^{\top}\boldsymbol\beta_{ ( M_0 + 1 ) } - \cdots - \pi_{ 1 d }\pi_{ ( M_{0} + M_{1} ) d }\boldsymbol Z_{( M_{0} + M_{1} )d}^{\top}\boldsymbol\beta_{( M_{0} + M_{1} )} \right] \\
&\vdots\\
+& \left[ - \pi_{ 1 d }\pi_{ ( \sum_{j=0}^{k-1}M_j + 1 ) d }\boldsymbol Z_{( \sum_{j=0}^{k-1}M_j + 1 )d}^{\top}\boldsymbol\beta_{ ( \sum_{j=0}^{k-1}M_j + 1 ) } - \cdots - \pi_{ 1 d }\pi_{ ( \sum_{j=0}^{k}M_j )d }\boldsymbol Z_{( \sum_{j=0}^{k}M_j )d}^{\top}\boldsymbol\beta_{ ( \sum_{j=0}^{k}M_j ) } \right]. \\
\end{align*} 
\normalsize
Hence, in the similar way, we have 
\begin{align*}
& E( Y_{i, d}( A_{ i ( \sum_{j=0}^{k} M_{ j }  ) d } - \pi_{ ( \sum_{j=0}^{k} M_{ j }  ) d } ) ) \\
=  & \left[ -\pi_{ ( \sum_{j=0}^{k}M_j ) d }\pi_{ 1 d }\boldsymbol Z_{1d}^{\top}\boldsymbol\beta_1- \cdots - \pi_{ ( \sum_{j=0}^{k}M_j ) d }\pi_{ M_{0} d }\boldsymbol Z_{ M_0 d}^{\top}\boldsymbol\beta_{M_0} \right] \\
+& \left[  - \pi_{ ( \sum_{j=0}^{k}M_j ) d }\pi_{ ( M_0 + 1 ) d } \boldsymbol Z_{( M_0 + 1 ) d}^{\top}\boldsymbol\beta_{ ( M_0 + 1 ) } - \cdots - \pi_{ ( \sum_{j=0}^{k}M_j ) d }\pi_{ ( M_{0} + M_{1} ) d }\boldsymbol Z_{( M_{0} + M_{1} )d}^{\top}\boldsymbol\beta_{( M_{0} + M_{1} )} \right] \\
&\vdots\\
+& \left[ -\pi_{ ( \sum_{j=0}^{k}M_j ) d }\pi_{ ( \sum_{j=0}^{k-1}M_j + 1 ) d } \boldsymbol Z_{( \sum_{j=0}^{k-1}M_j + 1 )d}^{\top}\boldsymbol\beta_{ ( \sum_{j=0}^{k-1}M_j + 1 ) } - \cdots + \pi_{ ( \sum_{j=0}^{k}M_j ) d }( 1- \pi_{ ( \sum_{j=0}^{k}M_j )d } )\boldsymbol Z_{( \sum_{j=0}^{k}M_j ) d}^{\top}\boldsymbol\beta_{ ( \sum_{j=0}^{k}M_j ) } \right]. \\
\end{align*}
\normalsize
The expectation part for the first summation of equation (\ref{theta_tilde_aa}) can be expressed by
\begin{align*}
& E\left( I_{id} \boldsymbol X_{id} \boldsymbol X_{id}^{\top} \right)\\
&E\left( I_{id} \right)E\left( \boldsymbol X_{id} \boldsymbol X_{id}^{\top} \right)\\
=& \tau_{ d } E\left\lbrace \begin{bmatrix}  
& \boldsymbol B_{d} \\
&  ( A_{ i 1 d } - \pi_{ 1 d } ) \boldsymbol Z_{1d}  \\
& \vdots  \\
& ( A_{ i \sum_{ j=0 }^{ k } M_{ j } d } - \pi_{ \sum_{ j=0 }^{ k } M_{ j } d } ) \boldsymbol Z_{ ( \sum_{ j=0 }^{ k } M_{ j } ) d}  \\
\end{bmatrix} 
\begin{bmatrix}  
& \boldsymbol B_{d}^{\top}, ( A_{ i 1 d } - \pi_{ 1 d } ) \boldsymbol Z_{1d}^{\top} , \cdots, ( A_{ i \sum_{j=0}^{k}M_{j} d } - \pi_{ \sum_{j=0}^{k} M_{j} d } ) \boldsymbol Z_{ ( \sum_{j=0}^{k}M_{j} ) d}^{\top}  \\
\end{bmatrix} \right\rbrace \\
=& \tau_{ d } E \begin{bmatrix}
\boldsymbol B_{d} \boldsymbol B_{d}^{\top} & 
\cdots &
(A_{i \sum_{j=0}^{k}M_j d}-\pi_{\sum_{j=0}^{k}M_j d})\boldsymbol B_{d}\boldsymbol Z_{ (\sum_{j=0}^{k}M_j ) d}^{\top}
\\
\vdots & 
\ddots &
\vdots
\\
(A_{i \sum_{j=0}^{k}M_j d}-\pi_{ \sum_{j=0}^{k}M_j d})\boldsymbol Z_{ ( \sum_{j=0}^{k}M_j ) d} \boldsymbol B_{d}^{\top} &
\cdots &
(A_{i \sum_{j=0}^{k}M_j d}-\pi_{ \sum_{j=0}^{k}M_j d})(A_{i \sum_{j=0}^{k}M_j d}-\pi_{ \sum_{j=0}^{k}M_j d})\boldsymbol Z_{ ( \sum_{j=0}^{k}M_j ) d}\boldsymbol Z_{ ( \sum_{j=0}^{k}M_j )  d}^{\top}
\end{bmatrix}
\\
=&\tau_{d} \begin{bmatrix}
\boldsymbol B_{d} \boldsymbol B_{d}^{\top} & 
0 & 
\cdots & 
0
\\
0 & 
\pi_{1d}(1-\pi_{1d})\boldsymbol Z_{1d} \boldsymbol Z_{1d}^{\top} &
\cdots &
- \pi_{1d}\pi_{ ( \sum_{j=0}^{k}M_j ) d}\boldsymbol Z_{1d} \boldsymbol Z_{ ( \sum_{j=0}^{k}M_j ) d}^{\top}
\\ 
\vdots & 
\vdots &
\ddots &
\vdots
\\
0 & 
- \pi_{ ( \sum_{j=0}^{k}M_j ) d}\pi_{1d}\boldsymbol Z_{ ( \sum_{j=0}^{k}M_j )  d} \boldsymbol Z_{1 d}^{\top} & 
\cdots & 
 \pi_{ ( \sum_{j=0}^{k}M_j ) d}(1-\pi_{ ( \sum_{j=0}^{k}M_j ) d})\boldsymbol Z_{( \sum_{j=0}^{k}M_j )d} \boldsymbol Z_{( \sum_{j=0}^{k}M_j )d}^{\top}
\\
\end{bmatrix}.
\end{align*}
\normalsize
Thus the estimator of equation (\ref{theta_tilde_aa}) is $\tilde{\boldsymbol\theta}$=$(\tilde{\boldsymbol\alpha}^{\top}, \tilde{\boldsymbol\beta}^{\top})^{\top}$, where 
$\tilde{\boldsymbol\alpha}$ is the first $q$ elements of $\tilde{\boldsymbol\theta}$ 
while 
$\tilde{\boldsymbol\beta}$ is the rest $\sum_{m=1}^{M}p_m$ elements of $\tilde{\boldsymbol\theta}$, i.e.
\begin{equation*}
\tilde{\boldsymbol\alpha}=\left( \sum_{d=1}^D\tau_{d}\boldsymbol B_{d}\boldsymbol B_{d}^{\top} \right)^{-1} \sum_{d=1}^D\tau_{d}\boldsymbol B_{d}^{\top}\boldsymbol\alpha\boldsymbol B_{d}
\end{equation*}
and
\begin{align*}
\tilde{\boldsymbol\beta}&=(\tilde{\boldsymbol\beta}_{1}^{\top},\ldots,\tilde{\boldsymbol\beta}_{M}^{\top})^{\top}\\
=\sum_{d=1}^D\tau_{d} &\begin{bmatrix}
\pi_{1d}(1-\pi_{1d})\boldsymbol Z_{1 d} \boldsymbol Z_{1 d}^{\top} &
\cdots &
 - \pi_{1d}\pi_{ ( \sum_{j=0}^{k}M_j ) d}\boldsymbol Z_{1 d} \boldsymbol Z_{ ( \sum_{j=0}^{k}M_j ) d}^{\top}
\\ 
\vdots & 
\ddots &
\vdots
\\
 - \pi_{ ( \sum_{j=0}^{k}M_j ) d}\pi_{1d}\boldsymbol Z_{ ( \sum_{j=0}^{k}M_j ) d} \boldsymbol Z_{1 d}^{\top} &  
\cdots & 
 \pi_{ ( \sum_{j=0}^{k}M_j ) d}(1-\pi_{ ( \sum_{j=0}^{k}M_j ) d})\boldsymbol Z_{ ( \sum_{j=0}^{k}M_j ) d} \boldsymbol Z_{ ( \sum_{j=0}^{k}M_j ) d}^{\top}
\\
\end{bmatrix}^{-1}\\
\sum_{d=1}^D\tau_{d} &\begin{bmatrix}
\left[ \pi_{1d}(1-\pi_{1d})\boldsymbol Z_{1 d}^{\top}\boldsymbol\beta_1  - \cdots - \pi_{1d}\pi_{ ( \sum_{j=0}^{k}M_j ) d}\boldsymbol Z_{( \sum_{j=0}^{k}M_j ) d}^{\top}\boldsymbol\beta_{ ( \sum_{j=0}^{k}M_j ) }  \right] \boldsymbol Z_{1 d}\\
\vdots\\
 \left[ -\pi_{1d}\pi_{ ( \sum_{j=0}^{k}M_j ) d}\boldsymbol Z_{1 d}^{\top}\boldsymbol\beta_1  - \cdots + \pi_{ ( \sum_{j=0}^{k}M_j ) d}(1-\pi_{ ( \sum_{j=0}^{k}M_j ) d})\boldsymbol Z_{ ( \sum_{j=0}^{k}M_j ) d}^{\top}\boldsymbol\beta_{ ( \sum_{j=0}^{k}M_j ) }  \right] \boldsymbol Z_{ ( \sum_{j=0}^{k}M_j ) d}\\
\end{bmatrix}.
\end{align*}

Next, we give detailed expressions for $\sqrt{N}(\hat{\boldsymbol\theta}-\tilde{\boldsymbol\theta})$ by substituting equation (\ref{theta_hat}) for the estimator $\hat{\boldsymbol\theta}$
, i.e.
\begin{align*}
&\sqrt{N}(\hat{\boldsymbol\theta}-\tilde{\boldsymbol\theta})\\
=&\sqrt{N}\left\lbrace \left[ \dfrac{1}{N}\sum_{i=1}^{N}\sum_{d=1}^{D}I_{id}\boldsymbol X_{id} \boldsymbol X_{id}^{\top} \right]^{-1}\dfrac{1}{N}\sum_{i=1}^{N}\sum_{d=1}^{D}I_{id}Y_{i, d}\boldsymbol X_{id} - \tilde{\boldsymbol\theta} \right\rbrace\\
=&\sqrt{N}\left\lbrace \left[ \dfrac{1}{N}\sum_{i=1}^{N}\sum_{d=1}^{D}I_{id}\boldsymbol X_{id} \boldsymbol X_{id}^{\top} \right]^{-1} \dfrac{1}{N}\sum_{i=1}^{N}\sum_{d=1}^{D}\left[ I_{id}Y_{i, d}\boldsymbol X_{id} - I_{id}\boldsymbol X_{id} \boldsymbol X_{id}^{\top} \tilde{\boldsymbol\theta} \right] \right\rbrace\\
=&\sqrt{N}\left\lbrace \left[ \sum_{d=1}^{D}E( I_{id}\boldsymbol X_{id} \boldsymbol X_{id}^{\top}) \right]^{-1} \dfrac{1}{N}\sum_{i=1}^{N}\sum_{d=1}^{D}\left[ I_{id}\boldsymbol X_{id} \tilde{\epsilon}_{id} \right] \right\rbrace + o_p(1),\\
\end{align*}
where $o_p(1)\xrightarrow{N\rightarrow\infty}0$, $\tilde{\epsilon}_{id}$=$Y_{i, d}$ - $\boldsymbol B_{d}^{\top}\tilde{\boldsymbol\alpha}-(A_{i1d}-\pi_{1d})\boldsymbol Z_{1d}^{\top}\tilde{\boldsymbol\beta}_1- \cdots -(A_{iMd}-\pi_{Md})\boldsymbol Z_{Md}^{\top}\tilde{\boldsymbol\beta}_M$ and $E( I_{id}\boldsymbol X_{id} \tilde{\epsilon}_{id})$=0.
Therefore, the asymptotic covariance matrix of $\hat{\boldsymbol\theta}$ (\ref{Sigma_theta}) can be derived by
\begin{align*}
\boldsymbol\Sigma_{\boldsymbol\theta}
=&\sqrt{N}\left[ \sum_{d=1}^{D}E( I_{id}\boldsymbol X_{id} \boldsymbol X_{id}^{\top}) \right]^{-1} 
E\left( 
\dfrac{1}{N}\sum_{i=1}^{N}\sum_{d=1}^{D}\left[ I_{id} \tilde{\epsilon}_{id} \boldsymbol X_{id} \right]   
\dfrac{1}{N}\sum_{i=1}^{N}\sum_{d=1}^{D}\left[ I_{id} \tilde{\epsilon}_{id} \boldsymbol X_{id}^{\top} \right] 
\right)
 \sqrt{N}\left[ \sum_{d=1}^{D}E( I_{id}\boldsymbol X_{id} \boldsymbol X_{id}^{\top}) \right]^{-1} 
  \\
=&\sqrt{N}\left[ \sum_{d=1}^{D}E( I_{id}\boldsymbol X_{id} \boldsymbol X_{id}^{\top}) \right]^{-1} 
\dfrac{1}{N}E\left( 
\sum_{d=1}^{D} I_{id} \tilde{\epsilon}_{id} \boldsymbol X_{id}  
\sum_{d=1}^{D}I_{id} \tilde{\epsilon}_{id} \boldsymbol X_{id}^{\top} 
\right)
 \sqrt{N}\left[ \sum_{d=1}^{D}E( I_{id}\boldsymbol X_{id} \boldsymbol X_{id}^{\top}) \right]^{-1}\\
 =&\left[ \sum_{d=1}^{D}E( I_{id}\boldsymbol X_{id} \boldsymbol X_{id}^{\top}) \right]^{-1} 
E\left( 
\sum_{d=1}^{D} I_{id} \tilde{\epsilon}_{id} \boldsymbol X_{id}  
\sum_{d=1}^{D}I_{id} \tilde{\epsilon}_{id} \boldsymbol X_{id}^{\top} 
\right)
\left[ \sum_{d=1}^{D}E( I_{id}\boldsymbol X_{id} \boldsymbol X_{id}^{\top}) \right]^{-1}.
\end{align*}
We have 
\begin{align*}
&E(\sum_{d=1}^{D} I_{id}\tilde{\epsilon}_{id}\boldsymbol X_{id}  \sum_{d=1}^{D} I_{id}\tilde{\epsilon}_{id}\boldsymbol X_{id}^{\top} )\\
=&E(\sum_{d=1}^{D} I_{id}\tilde{\epsilon}_{id}^2\boldsymbol X_{id}\boldsymbol X_{id}^{\top} + \sum_{(d)\neq(d)^\prime}^{(D)} I_{i(d)}I_{i(d)^\prime}\tilde{\epsilon}_{i(d)} \tilde{\epsilon}_{i(d)^\prime} \boldsymbol X_{i(d)}\boldsymbol X_{i(d)^\prime}^{\top})\\
=&\sum_{d=1}^{D}\sigma_{d}^2 \tau_{d}\begin{bmatrix}
 \boldsymbol B_{d} \boldsymbol B_{d}^{\top} & 
0 & 
\cdots & 
0
\\
0 & 
 \pi_{1d}(1-\pi_{1d})\boldsymbol Z_{1 d} \boldsymbol Z_{1 d}^{\top} &
\cdots &
- \pi_{1d}\pi_{ ( \sum_{j=0}^{k}M_j  ) d}\boldsymbol Z_{1 d} \boldsymbol Z_{( \sum_{j=0}^{k}M_j  ) d}^{\top} 
\\ 
\vdots & 
\vdots &
\ddots &
\vdots
\\
0 & 
- \pi_{ ( \sum_{j=0}^{k}M_j  ) dt}\pi_{1d}\boldsymbol Z_{( \sum_{j=0}^{k}M_j  ) d} \boldsymbol Z_{1 d}^{\top}  & 
\cdots & 
 \pi_{ ( \sum_{j=0}^{k}M_j ) d}(1-\pi_{ ( \sum_{j=0}^{k}M_j  ) d})\boldsymbol Z_{( \sum_{j=0}^{k}M_j  ) d}\boldsymbol Z_{( \sum_{j=0}^{k}M_j  ) d}^{\top} 
\\
\end{bmatrix}\\
&+ \sum_{(d)\neq (d)^\prime}^{(D)} \sigma_{(d)(d)^\prime}\tau_{d}\tau_{(d)^\prime} \begin{bmatrix}
\boldsymbol B_{d} \boldsymbol B_{(d)^\prime}^{\top} & 
0 & 
\cdots & 
0
\\
0 & 
0 &
\cdots &
0
\\ 
\vdots & 
\vdots &
\ddots &
\vdots
\\
0 & 
0 & 
\cdots & 
0
\\
\end{bmatrix}\\
\approx&\bar{\sigma}^2\sum_{d=1}^{D} \tau_{d}\begin{bmatrix}
 \boldsymbol B_{d} \boldsymbol B_{d}^{\top} & 
0 & 
\cdots & 
0
\\
0 & 
 \pi_{1d}(1-\pi_{1d})\boldsymbol Z_{1 d} \boldsymbol Z_{1 d}^{\top} &
\cdots &
- \pi_{1d}\pi_{ ( \sum_{j=0}^{k}M_j  ) d}\boldsymbol Z_{1 d} \boldsymbol Z_{( \sum_{j=0}^{k}M_j  ) d}^{\top} 
\\ 
\vdots & 
\vdots &
\ddots &
\vdots
\\
0 & 
- \pi_{ ( \sum_{j=0}^{k}M_j  ) dt}\pi_{1d}\boldsymbol Z_{( \sum_{j=0}^{k}M_j  ) d} \boldsymbol Z_{1 d}^{\top}  & 
\cdots & 
 \pi_{ ( \sum_{j=0}^{k}M_j ) d}(1-\pi_{ ( \sum_{j=0}^{k}M_j  ) d})\boldsymbol Z_{( \sum_{j=0}^{k}M_j  ) d}\boldsymbol Z_{( \sum_{j=0}^{k}M_j  ) d}^{\top} 
\\
\end{bmatrix}\\
&+ \sum_{(d)\neq (d)^\prime}^{(D)} \sigma_{(d)(d)^\prime}\tau_{d}\tau_{(d)^\prime} \begin{bmatrix}
\boldsymbol B_{d} \boldsymbol B_{(d)^\prime}^{\top} & 
0 & 
\cdots & 
0
\\
0 & 
0 &
\cdots &
0
\\ 
\vdots & 
\vdots &
\ddots &
\vdots
\\
0 & 
0 & 
\cdots & 
0
\\
\end{bmatrix},
\end{align*}
where 
$\bar{\sigma}^2$=$\sum_{d=1}^{D}\text{Var}(\epsilon_{id})/D$.
Thus, assuming constant variance $\text{Var}(\epsilon_{id})$ over $d$, i.e. $\bar{\sigma}^2$=$\sigma^2$, $ \sigma_{(d)(d)^\prime}=0$, or $\epsilon_{id}$ and $\epsilon_{i(d)^\prime}$ are independent, where $(d)\neq (d)^\prime$=$1,\ldots, D$, the asymptotic covariance matrix $\boldsymbol\Sigma_{\boldsymbol\beta}$ is the lower right $\sum_{m=1}^{M}p_m\times\sum_{m=1}^{M}p_m$ block of the asymptotic covariance matrix $\boldsymbol\Sigma_{\boldsymbol\theta}$,  can be defined by
\begin{align*}
&\boldsymbol\Sigma_{\boldsymbol\beta}\\
\approx &\sum_{d=1}^D\tau_{d}
\begin{bmatrix}
 \pi_{1d}(1-\pi_{1d})\boldsymbol Z_{1d} \boldsymbol Z_{1d}^{\top} &
\cdots &
 -\pi_{1d}\pi_{ ( \sum_{j=0}^{k}M_j ) d}\boldsymbol Z_{1d} \boldsymbol Z_{( \sum_{j=0}^{k}M_j  ) d}^{\top} 
\\ 
\vdots & 
\ddots &
\vdots
\\
 - \pi_{ ( \sum_{j=0}^{k}M_j ) d}\pi_{1d}\boldsymbol Z_{( \sum_{j=0}^{k}M_j  ) d} \boldsymbol Z_{1d}^{\top} &  
\cdots & 
 \pi_{ ( \sum_{j=0}^{k}M_j ) d}(1-\pi_{ ( \sum_{j=0}^{k}M_j ) d})\boldsymbol Z_{( \sum_{j=0}^{k}M_j  ) d} \boldsymbol Z_{( \sum_{j=0}^{k}M_j  ) d}^{\top}
\\
\end{bmatrix}^{-1}\\
\bar{\sigma}^2&\sum_{d=1}^D\tau_{d}\begin{bmatrix}
 \pi_{1d}(1-\pi_{1d})\boldsymbol Z_{1d} \boldsymbol Z_{1d}^{\top} &
\cdots &
 - \pi_{1d}\pi_{ ( \sum_{j=0}^{k}M_j ) d}\boldsymbol Z_{1d} \boldsymbol Z_{( \sum_{j=0}^{k}M_j  ) d}^{\top}
\\ 
\vdots & 
\ddots &
\vdots
\\
 - \pi_{ ( \sum_{j=0}^{k}M_j ) d}\pi_{1d}\boldsymbol Z_{( \sum_{j=0}^{k}M_j  ) d} \boldsymbol Z_{1d}^{\top} &  
\cdots & 
 \pi_{ ( \sum_{j=0}^{k}M_j ) d}(1-\pi_{ ( \sum_{j=0}^{k}M_j ) d})\boldsymbol Z_{( \sum_{j=0}^{k}M_j  ) d} \boldsymbol Z_{( \sum_{j=0}^{k}M_j  ) d}^{\top}
\\
\end{bmatrix}\\
&\sum_{d=1}^D\tau_{d}\begin{bmatrix}
 \pi_{1d}(1-\pi_{1d})\boldsymbol Z_{1d} \boldsymbol Z_{1d}^{\top} &
\cdots &
 - \pi_{1d}\pi_{ ( \sum_{j=0}^{k}M_j ) d}\boldsymbol Z_{1d} \boldsymbol Z_{( \sum_{j=0}^{k}M_j  ) d}^{\top}
\\ 
\vdots & 
\ddots &
\vdots
\\
 - \pi_{ ( \sum_{j=0}^{k}M_j ) d}\pi_{1d}\boldsymbol Z_{( \sum_{j=0}^{k}M_j  ) d} \boldsymbol Z_{1d}^{\top} &  
\cdots & 
 \pi_{ ( \sum_{j=0}^{k}M_j ) d}(1-\pi_{ ( \sum_{j=0}^{k}M_j ) d})\boldsymbol Z_{( \sum_{j=0}^{k}M_j  ) d} \boldsymbol Z_{( \sum_{j=0}^{k}M_j  ) d}^{\top}
\\
\end{bmatrix}^{-1}\\
=\bar{\sigma}^2&\sum_{d=1}^D\tau_{d}\begin{bmatrix}
 \pi_{1d}(1-\pi_{1d})\boldsymbol Z_{1d} \boldsymbol Z_{1d}^{\top} &
\cdots &
 - \pi_{1d}\pi_{ ( \sum_{j=0}^{k}M_j ) d}\boldsymbol Z_{1d} \boldsymbol Z_{( \sum_{j=0}^{k}M_j  ) d}^{\top}
\\ 
\vdots & 
\ddots &
\vdots
\\
 - \pi_{ ( \sum_{j=0}^{k}M_j ) d}\pi_{1d}\boldsymbol Z_{( \sum_{j=0}^{k}M_j  ) d} \boldsymbol Z_{1d}^{\top} &  
\cdots & 
 \pi_{ ( \sum_{j=0}^{k}M_j ) d}(1-\pi_{ ( \sum_{j=0}^{k}M_j ) d})\boldsymbol Z_{( \sum_{j=0}^{k}M_j  ) d} \boldsymbol Z_{( \sum_{j=0}^{k}M_j  ) d}^{\top}
\\
\end{bmatrix}^{-1}.
\end{align*}
\normalsize

\section{Proof of Lemma 2.1}\label{app: ThePro}%
\noindent \textbf{Proof:}  The proof of consistency requires the result of the strong law of large numbers, such that  
$\hat{\boldsymbol\theta}\rightarrow\tilde{\boldsymbol\theta}$, almost surely, and uniformly for $\boldsymbol\theta\in\boldsymbol\Theta$ 
as $N\rightarrow\infty$ and $\tilde{\boldsymbol\theta}$ being the unique minimum value of $E(SEC(\boldsymbol\theta))$. 
To prove the asymptotic normality result, by the central limit theorem, $\sqrt{N}(\hat{\boldsymbol\theta}-\tilde{\boldsymbol\theta})$ converges in distribution to $\text{Normal}(0,\boldsymbol\Sigma_{\boldsymbol\theta})$.

Based on the proof Lemma \ref{theo1}, we conclude that the asymptotic distribution of $\sqrt{N}(\hat{\boldsymbol\beta}-\tilde{\boldsymbol\beta})$ converges to $\text{Normal}(0,\boldsymbol\Sigma_{\boldsymbol\beta})$.

\section{Proof of Corollary 2.1.1}\label{app: DistrTest}
Based on the equations (\ref{e_hat_id}) to (\ref{H_i}) defined in Section \ref{test}, and the covariance estimator of GEE regression coefficient derivation in \cite{Mancl_DeRouen_2001}, we have an approximation, i.e.
\begin{equation}\label{E_ee_hat_i}
E(\hat{\boldsymbol e}_{i} \hat{\boldsymbol e}_{i}^{\top}) \approx (\boldsymbol I_{D\times D} - \boldsymbol H_i)\text{COV}(\boldsymbol Y_i)(\boldsymbol I_{D\times D} - \boldsymbol H_i)^{\top} 
\end{equation}
and the covariance estimator of $\hat{\boldsymbol\theta}$, i.e.
\begin{equation}\label{Sigma_theta_hat}
\hat{\boldsymbol\Sigma}_{\boldsymbol\theta} = 
\left( \sum_{i=1}^N \boldsymbol X_{i} \boldsymbol X_{i}^{\top}/N \right)^{-1}
\left( \sum_{i=1}^N \boldsymbol X_{i} (\boldsymbol I_{D\times D} - \boldsymbol H_i)^{-1} \hat{\boldsymbol e}_{i} \hat{\boldsymbol e}_{i}^{\top} (\boldsymbol I_{D\times D} - \boldsymbol H_i)^{-1} \boldsymbol X_{i}^{\top} \right)
\left( \sum_{i=1}^N \boldsymbol X_{i} \boldsymbol X_{i}^{\top}/N \right)^{-1}.
\end{equation}
Based on equation (\ref{E_ee_hat_i}), assuming the second summation of equation (\ref{Sigma_theta_hat}) can be approximated by
\begin{align*}
&\sum_{i=1}^N \boldsymbol X_{i} (\boldsymbol I_{D\times D} - \boldsymbol H_i)^{-1} \hat{\boldsymbol e}_{i}\hat{\boldsymbol e}_{i}^{\top} (\boldsymbol I_{D\times D} - \boldsymbol H_i)^{-1} \boldsymbol X_{i}^{\top}\\
\approx & \sum_{i=1}^N \boldsymbol X_{i} \text{COV}(\boldsymbol Y_i) \boldsymbol X_{i}^{\top}\\
\approx & \sum_{i=1}^N \boldsymbol X_{i} \boldsymbol \epsilon_{i} \boldsymbol \epsilon_{i}^{\top} \boldsymbol X_{i}^{\top},
\end{align*}
where
\begin{align*}
\boldsymbol X_{i} \boldsymbol \epsilon_{i}  =
\begin{bmatrix}
\sum_{d=1}^{D}I_{id}  \epsilon_{id} \\
\vdots\\
\sum_{d=1}^{D} I_{id} ( d - 1 )^{q-1} \epsilon_{id} \\
\sum_{d=1}^{D} I_{id} ( A_{i1d} - \pi_{1d} ) \epsilon_{id} \\
\vdots\\
\sum_{d=1}^{D} I_{id} ( A_{i1d} - \pi_{1d} ) ( d - 1 )^{p_1-1} \epsilon_{id} \\
\vdots\\
\sum_{d=1}^{D} I_{id} ( A_{iM_{0}d} - \pi_{M_{0}d} ) \epsilon_{id} \\
\vdots\\
\sum_{d=1}^{D} I_{id} ( A_{iM_{0}d} - \pi_{M_{0}d} ) ( d - 1 )^{p_{M_{0}}-1} \epsilon_{id} \\
\vdots\\
\sum_{d=1}^{D} I_{id} ( A_{i\sum_{j=0}^{k}M_{j}d} - \pi_{\sum_{j=0}^{k}M_{j}d} ) \epsilon_{id} \\
\vdots\\
\sum_{d=1}^{D} I_{id} ( A_{i\sum_{j=0}^{k}M_{j}d} - \pi_{\sum_{j=0}^{k}M_{j}d} ) ( d - 1 )^{p_{ \sum_{j=0}^{k}M_{j} }-1} \epsilon_{id} \\
\end{bmatrix}
\end{align*}
with dimension $\left( q + \sum_{m=1}^M p_m \right) \times 1$. Hence, we can have $\hat{\boldsymbol\Sigma}_{\boldsymbol\beta\boldsymbol\beta}$ approximated by 
\begin{align*}
\boldsymbol U
\left( \sum_{i=1}^N \boldsymbol X_{i} \boldsymbol X_{i}^{\top}/N \right)^{-1}
\left( \sum_{i=1}^N \boldsymbol X_{i} \boldsymbol \epsilon_{i} \boldsymbol \epsilon_{i}^{\top} \boldsymbol X_{i}^{\top} \right)
\left( \sum_{i=1}^N \boldsymbol X_{i} \boldsymbol X_{i}^{\top}/N \right)^{-1}
\boldsymbol U^{\top},
\end{align*}
where $\boldsymbol U$ is a rectangular matrix with dimension $\left( \sum_{m=1}^M p_m \right) \times \left( q + \sum_{m=1}^M p_m \right)$,  i.e. $U = \left[ \boldsymbol U_0, \boldsymbol U_1 \right]$, where $\boldsymbol U_0$ is a zeros matrix with dimension $\sum_{m=1}^M p_m \times q$ while $\boldsymbol U_1$ is an identity matrix with dimension $\sum_{m=1}^M p_m \times \sum_{m=1}^M p_m$ .
Assuming $\boldsymbol X_{i}$ are given, but not $\boldsymbol\epsilon_i$ and hence $\boldsymbol X_{i} \boldsymbol \epsilon_{i}$ are random, for $i=1,\ldots N$. Therefore, the degrees of freedom for $\boldsymbol X_{i} \boldsymbol \epsilon_{i}$ can be $N$ when assuming no restrictions on $\boldsymbol X_{i} \boldsymbol \epsilon_{i}$. 
Thus, when $N$ is small, the test statistic follows a Hotelling's $T^2_{\sum_{m=1}^{M}p_m, N}$ distribution.

\subsection{Precision-Based Sample Size Calculation}\label{precisionsamplesize}
Based on the test statistic of \cite{Liao_etal_2016} with Hotelling's $T^2$ distribution in equation (\ref{N_F_N_q_1_P}), which assumes a small sample size $N$,
we define the coverage probability or confidence level ($1-\alpha$) in terms of sample size $N$ and precision $\hat{\boldsymbol\beta} - \tilde{\boldsymbol\beta}$ for estimator $\hat{\boldsymbol\beta}$,
\begin{align*}
&\text{Pr}\left(  0 < ( \hat{\boldsymbol\beta} - \tilde{\boldsymbol\beta} )^{\top}\hat{\boldsymbol\Sigma}_{\boldsymbol\beta}^{-1}( \hat{\boldsymbol\beta} - \tilde{\boldsymbol\beta} )  <  \dfrac{ \sum_{m=1}^{M}p_m(N-q-1)}{N(N-q- \sum_{m=1}^{M}p_m)} f_{ \sum_{m=1}^{M}p_m, N-q- \sum_{m=1}^{M}p_m, \alpha} \right)=1-\alpha,
\end{align*}
assuming $N>q+ \sum_{m=1}^{M}p_m$. 
Let $\boldsymbol\beta^{\ast} - \tilde{\boldsymbol\beta}$ denote the pre-specified desired precision and $\boldsymbol\Sigma^{\ast}_{\boldsymbol\beta}$ denote the covariance matrix for $\hat{\boldsymbol\beta}$  at $\hat{\boldsymbol\beta}=\boldsymbol\beta^{\ast}$.
Hence $N$ can be calculated by solving the following equation,
\begin{equation}\label{N_F_N_q_1_C}
( \boldsymbol\beta^{\ast} - \tilde{\boldsymbol\beta} )^{\top}(\boldsymbol\Sigma^{\ast}_{\boldsymbol\beta})^{-1}( \boldsymbol\beta^{\ast} - \tilde{\boldsymbol\beta} ) >( \hat{\boldsymbol\beta}- \tilde{\boldsymbol\beta} )^{\top}\hat{\boldsymbol\Sigma}_{\boldsymbol\beta}^{-1}( \hat{\boldsymbol\beta} - \tilde{\boldsymbol\beta} )
\sim\dfrac{ \sum_{m=1}^{M}p_m(N-q-1)}{N(N-q- \sum_{m=1}^{M}p_m)}f_{ \sum_{m=1}^{M}p_m, N-q- \sum_{m=1}^{M}p_m, \alpha}.
\end{equation}
In other words, $N$ is computed by limiting the upper bound of the random quantity $( \hat{\boldsymbol\beta}- \tilde{\boldsymbol\beta} )^{\top}\hat{\boldsymbol\Sigma}_{\boldsymbol\beta}^{-1}( \hat{\boldsymbol\beta} - \tilde{\boldsymbol\beta} )$ not higher than a pre-specified value, i.e. $( \boldsymbol\beta^{\ast} - \tilde{\boldsymbol\beta} )^{\top}(\boldsymbol\Sigma^{\ast}_{\boldsymbol\beta})^{-1}( \boldsymbol\beta^{\ast} - \tilde{\boldsymbol\beta} ) $.
Note that $N$ is varied only due to $\boldsymbol\beta^{\ast}-\tilde{\boldsymbol\beta} $ and everything else is fixed in equation (\ref{N_F_N_q_1_C}).
Under different test statistics in equations (\ref{N_chi_P}) and (\ref{N_F_N_P}), $N$ can be calculated respectively using
\begin{equation}\label{N_chi_C}
( \boldsymbol\beta^{\ast} - \tilde{\boldsymbol\beta} )^{\top}(\boldsymbol\Sigma^{\ast}_{\boldsymbol\beta})^{-1}( \boldsymbol\beta^{\ast} - \tilde{\boldsymbol\beta} )  > ( \hat{\boldsymbol\beta}- \tilde{\boldsymbol\beta} )^{\top}\hat{\boldsymbol\Sigma}_{\boldsymbol\beta}^{-1}( \hat{\boldsymbol\beta} - \tilde{\boldsymbol\beta} )
\sim\dfrac{ 1 }{ N }\chi_{ \sum_{m=1}^{M}p_m, \alpha}^2,
\end{equation}
and
\begin{equation}\label{N_F_N_C}
( \boldsymbol\beta^{\ast} - \tilde{\boldsymbol\beta} )^{\top}(\boldsymbol\Sigma^{\ast}_{\boldsymbol\beta})^{-1}( \boldsymbol\beta^{\ast} - \tilde{\boldsymbol\beta} )
> ( \hat{\boldsymbol\beta}- \tilde{\boldsymbol\beta} )^{\top}\hat{\boldsymbol\Sigma}_{\boldsymbol\beta}^{-1}( \hat{\boldsymbol\beta} - \tilde{\boldsymbol\beta} )
\sim\dfrac{ \sum_{m=1}^{M}p_m(N)}{N(N- \sum_{m=1}^{M}p_m + 1)}f_{ \sum_{m=1}^{M}p_m, N- \sum_{m=1}^{M}p_m + 1, \alpha}.
\end{equation}
The test statistic with $\chi^2$ distribution in equation (\ref{N_chi_C}) assumes a large $N$.
$N$ can be computed by selecting the minimum value that gives the estimated coverage probability not lower than its nominal value ($1-\alpha$).
Note that given the other quantities are the same, $N$ increases as the magnitude of $\mid\boldsymbol\beta^{\ast} - \tilde{\boldsymbol\beta}\mid$ decreases (i.e. estimate becomes more precise).

\section{DIAMANTE Study with Student Population Example: Precision-Based Sample Size Calculation}\label{s:realexample_precision}
\begin{figure}
\begin{center}
\begin{tabular}{c}
 \includegraphics[scale=0.5]{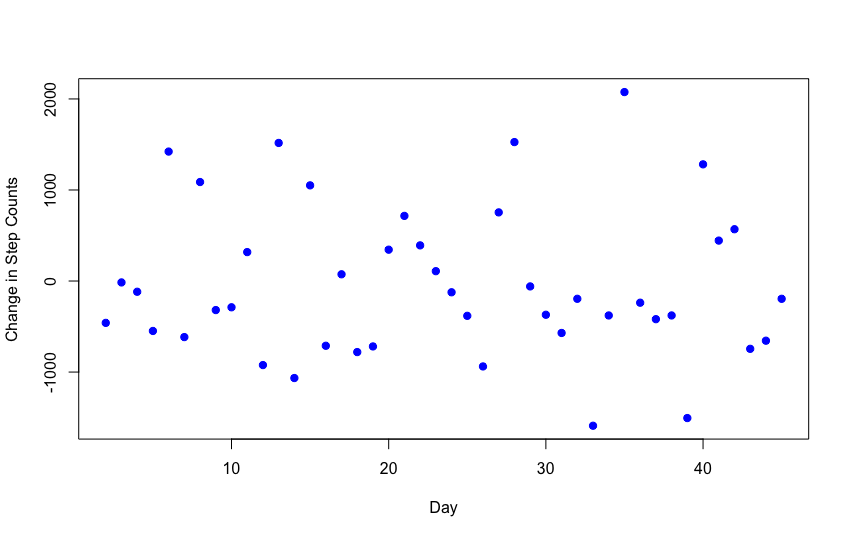} \\
\end{tabular}
\end{center}
\caption{
Average change of step counts from yesterday on each of days 2 to 45 
}
\label{fig: StepChange}
\end{figure}

The proximal effect sizes are likely to follow the constant trend based on the observations in Figure \ref{fig: StepChange}. 
Let the standardized precision of the initial and average proximal effect sizes denoted by $\Delta\boldsymbol\delta^0$ ($\Delta\boldsymbol\beta^0/\bar{\sigma}$) and $\Delta\bar{\boldsymbol\delta}^d$ ($\Delta\bar{\boldsymbol\beta}^d/\bar{\sigma}$) respectively, where we specify $\Delta\boldsymbol\delta^0$=$\Delta\bar{\boldsymbol\delta}^d$=$(0.073, 0.121, 0.108)^{\top}$. 
Note that unlike the power-based method, the precision-based method does not consider the possible proximal effect sizes, and it only considers the precision or margin of error of the proximal effect size of each intervention category.
Therefore, at $100\%$ availability and $44$ decision time points, using Hotelling's $T^2$ test statistic (\ref{N_F_N_q_1_C}) calculates a sample size of $86$ that achieves the desired precision for the proximal effect sizes with a probability of $95\%$.

\section{Simulation Study: Precision-Based Sample Size Calculation
}\label{s:sim_precision}
We define the precision of the initial proximal effect sizes $\delta_m(1, \hat{\boldsymbol\delta}_m) - \delta_m(1, \tilde{\boldsymbol\delta}_m)$ for $m$=1, 2 and 3, and $\delta_m(\lfloor 0.5D \rfloor+1, \hat{\boldsymbol\delta}_m) - \delta_m(\lfloor 0.5D \rfloor+1, \tilde{\boldsymbol\delta}_m)$ for $m=4$ to be $0.001$; 
the precision of the average standardized proximal effect sizes $\dfrac{1}{D}\sum_{d=1}^{D}\left(\delta_m(d, \hat{\boldsymbol\delta}_m) - \delta_m(d, \tilde{\boldsymbol\delta}_m)\right)$ for $m$=1, 2 and 3,  and \\
$\dfrac{1}{ D - \lfloor 0.5D \rfloor }\sum_{d=\lfloor 0.5D \rfloor+1}^{D}\left(\delta_m(d, \hat{\boldsymbol\delta}_m) - \delta_m(d, \tilde{\boldsymbol\delta}_m)\right)$ for $m=4$ to be $0.10$ and $0.06$.
The nominal coverage probability is $95\%$.
Note that $\hat{\cdot}$ represents an estimator based on the finite samples, $\tilde{\cdot}$ represents an estimator based on the infinite samples, $\lfloor\cdot\rfloor$ represents the floor function.

\subsection{Correctly specified working model}\label{s:correct model precision}
Table \ref{Table: T2} gives the calculated $N$, and the corresponding formulated and MC coverage probability ($CP$) estimates.
We observe that both the $CP$ estimates are similar. In general, when the working model is correctly specified, the proposed sample size calculator provides accurate $CP$. 

\begin{table}[h]
\centering
\caption{
The $N$ and $CP$ estimates
when the working model assumptions are correct. 
}
\label{Table: T2}
\begin{tabular}{rllrrrrrrr}
  \hline
 & \multicolumn{3}{c}{} &  \multicolumn{6}{c}{Precision of average standardized proximal effect size}  \\
 & Availability & Test Statistics & Decision Time Point & 0.10 & 0.06 & 0.10 & 0.06 & 0.10 & 0.06 \\ 
 & \multicolumn{3}{c}{} &  \multicolumn{2}{c}{$N$} & \multicolumn{2}{c}{Formulated $CP$}  & \multicolumn{2}{c}{Monte Carlo $CP$} \\
  \hline
& 100\% & $\chi^2_{\sum_{m}p_m}$                          & 180 &  47 & 132 & 0.95 & 0.95 & 0.94 & 0.96 \\ 
&            &                                                                   &  90 &  88 & 249 & 0.95 & 0.95 & 0.95 & 0.95 \\ 
&            & Hotelling's $T^2_{\sum_{m}p_m, N}$       & 180 &  59 & 143 & 0.95 & 0.95 & 0.95 & 0.95 \\ 
&            &                                                                   &  90 & 100 & 261 & 0.95 & 0.95 & 0.96 & 0.95 \\ 
&            & Hotelling's $T^2_{\sum_{m}p_m, N-q-1}$ & 180 &  59 & 143 & 0.95 & 0.95 & 0.96 & 0.95 \\ 
&            &                                                                   &  90 & 100 & 261 & 0.95 & 0.95 & 0.94 & 0.96 \\ 
& 70\%   & $\chi^2_{\sum_{m}p_m}$                          & 180 &  67 & 188 & 0.95 & 0.95 & 0.94 & 0.95 \\ 
&            &                                                                   &  90 & 126 & 356 & 0.95 & 0.95 & 0.95 & 0.95 \\ 
&            & Hotelling's $T^2_{\sum_{m}p_m, N}$       & 180 &  79 & 199 & 0.95 & 0.95 & 0.95 & 0.95 \\ 
&            &                                                                   &  90 & 138 & 368 & 0.95 & 0.95 & 0.96 & 0.96 \\ 
&            & Hotelling's $T^2_{\sum_{m}p_m, N-q-1}$ & 180 &  79 & 200 & 0.95 & 0.95 & 0.96 & 0.97 \\ 
&            &                                                                   &  90 & 138 & 368 & 0.95 & 0.95 & 0.94 & 0.95 \\ 
   \hline
\end{tabular}
\end{table}

\newpage

\subsection{Mis-specified trend for proximal effect}\label{s:mis trend precision}

In Table \ref{Table: CP_MisTrend}, we observe that the formulated $CP$ estimates are very close to the nominal $CP$, but the MC $CP$ estimates are much lower for the constant trend, slightly lower for the linear trend and slightly higher for the quadratic trend.

\begin{table}[h]
\centering
\caption{
The $N$ and $CP$ estimates
when the standardized proximal effect size trends are mis-specified.
}
\label{Table: CP_MisTrend}
\begin{tabular}{rlllrrrrrr}
  \hline
 & \multicolumn{3}{c}{} &  \multicolumn{6}{c}{Precision of average standardized proximal effect size}  \\
 &Trend & Availability & Test Statistics & 0.1 & 0.06 & 0.1 & 0.06 & 0.1 & 0.06 \\ 
 & \multicolumn{3}{c}{} &  \multicolumn{2}{c}{$N$} & \multicolumn{2}{c}{Formulated $CP$}  & \multicolumn{2}{c}{Monte Carlo $CP$} \\
  \hline
& Constant & 100\% & $\chi^2_{\sum_m p_m}$                          &  31 &  85 & 0.95 & 0.95 & 0.75 & 0.73 \\ 
&                &             & Hotelling's $T^2_{\sum_m p_m, N} $      &  37 &  91 & 0.95 & 0.95 & 0.69 & 0.69 \\ 
&                &             & Hotelling's $T^2_{\sum_m p_m, N-q-1}$ &  38 &  92 & 0.95 & 0.95 & 0.73 & 0.73 \\ 
&                & 70\%    & $\chi^2_{\sum_m p_m}$                          &  44 & 121 & 0.95 & 0.95 & 0.75 & 0.75 \\ 
&                &             & Hotelling's $T^2_{\sum_m p_m, N} $      &  50 & 128 & 0.95 & 0.95 & 0.71 & 0.75 \\ 
&                &             & Hotelling's $T^2_{\sum_m p_m, N-q-1}$ &  51 & 128 & 0.95 & 0.95 & 0.73 & 0.72 \\
\\
& Linear & 100\% & $\chi^2_{\sum_m p_m}$                         &  42 & 120 & 0.95 & 0.95 & 0.92 & 0.93 \\ 
&            &            & Hotelling's $T^2_{\sum_m p_m, N} $      &  54 & 132 & 0.95 & 0.95 & 0.92 & 0.94 \\ 
&            &            & Hotelling's $T^2_{\sum_m p_m, N-q-1}$ &  54 & 132 & 0.95 & 0.95 & 0.93 & 0.94 \\ 
&            & 70\%   & $\chi^2_{\sum_m p_m}$                          &  60 & 171 & 0.95 & 0.95 & 0.93 & 0.92 \\ 
&            &            & Hotelling's $T^2_{\sum_m p_m, N} $      &  71 & 183 & 0.95 & 0.95 & 0.92 & 0.92 \\ 
&            &            & Hotelling's $T^2_{\sum_m p_m, N-q-1}$ &  72 & 183 & 0.95 & 0.95 & 0.93 & 0.93 \\ 
\\
& Quadratic & 100\% & $\chi^2_{\sum_m p_m}$                          &  49 & 140 & 0.96 & 0.95 & 0.97 & 0.97 \\ 
&                 &             & Hotelling's $T^2_{\sum_m p_m, N} $      &  65 & 156 & 0.95 & 0.95 & 0.98 & 0.97 \\ 
&                 &             & Hotelling's $T^2_{\sum_m p_m, N-q-1}$ &  66 & 157 & 0.95 & 0.95 & 0.97 & 0.97 \\ 
&                 & 70\%    & $\chi^2_{\sum_m p_m}$                          &  69 & 199 & 0.95 & 0.95 & 0.95 & 0.94 \\ 
&                 &             & Hotelling's $T^2_{\sum_m p_m, N} $      &  85 & 216 & 0.95 & 0.95 & 0.97 & 0.97 \\ 
&                 &             & Hotelling's $T^2_{\sum_m p_m, N-q-1}$ &  86 & 216 & 0.95 & 0.95 & 0.98 & 0.97 \\
   \hline
\end{tabular}
\end{table}

\newpage

\subsection{Mis-specified number of intervention categories}\label{s:mis cat precision}
In Table  \ref{Table: CP_MisCat}, the formulated $CP$ estimates are very close to $95\%$, but because the $N$s are calculated assuming fewer categories, the MC $CP$ estimates are much lower than $95\%$. The MC $CP$ estimates performed worst when $M_0=1$ and $M_1=0$, as they are the furthest away from the true values (i.e. $M_0=3$ and $M_1=1$). 

\begin{table}[h]
\centering
\caption{
The $N$ and $CP$ estimates
when $M_0$ and $M_1$, and consequently $M$ are mis-specified.
}
\label{Table: CP_MisCat}
\begin{tabular}{rlllrrrrrr}
  \hline
 & \multicolumn{3}{c}{} &  \multicolumn{6}{c}{Precision of average standardized proximal effect size}  \\
 & $M_0$ and $M_1$ & Availability & Test Statistics & 0.1 & 0.06 & 0.1 & 0.06 & 0.1 & 0.06 \\ 
& \multicolumn{3}{c}{} &  \multicolumn{2}{c}{$N$} & \multicolumn{2}{c}{Formulated $CP$}  & \multicolumn{2}{c}{Monte Carlo $CP$} \\
  \hline
  & $M_0=1$ and $M_1=0$ & 100\% & $\chi^2_{\sum_m p_m}$                          & 13 &  36 & 0.95 & 0.95 & 0.18 & 0.18 \\ 
  &                                        &            & Hotelling's $T^2_{\sum_m p_m, N}$        &  17 &  40 & 0.95 & 0.95 & 0.10 & 0.15 \\ 
  &                                        &            & Hotelling's $T^2_{\sum_m p_m, N-q-1}$  &  18 &  40 & 0.96 & 0.95 & 0.14 & 0.15 \\ 
  &                                        & 70\%  & $\chi^2_{\sum_m p_m}$                            &  19 &  51 & 0.96 & 0.95 & 0.18 & 0.16 \\ 
  &                                        &           & Hotelling's $T^2_{\sum_m p_m, N}$          &  23 &  55 & 0.96 & 0.95 & 0.16 & 0.14 \\ 
  &                                        &           & Hotelling's $T^2_{\sum_m p_m, N-q-1}$   &  23 &  56 & 0.95 & 0.95 & 0.15 & 0.16 \\ 
  \\
  & $M_0=3$ and $M_1=0$ & 100\% & $\chi^2_{\sum_m p_m}$                         &  36 & 100 & 0.95 & 0.95 & 0.82 & 0.86 \\ 
  &                                        &            & Hotelling's $T^2_{\sum_m p_m, N}$       &  45 & 110 & 0.95 & 0.95 & 0.85 & 0.82 \\ 
  &                                        &            & Hotelling's $T^2_{\sum_m p_m, N-q-1}$ &  46 & 110 & 0.95 & 0.95 & 0.86 & 0.86 \\ 
  &                                        & 70\%   & $\chi^2_{\sum_m p_m}$                          &  52 & 143 & 0.95 & 0.95 & 0.82 & 0.85 \\ 
  &                                        &            & Hotelling's $T^2_{\sum_m p_m, N}$        &  61 & 152 & 0.95 & 0.95 & 0.86 & 0.83 \\ 
  &                                        &            & Hotelling's $T^2_{\sum_m p_m, N-q-1}$  &  61 & 153 & 0.95 & 0.95 & 0.84 & 0.86 \\
  \\
  & $M_0=4$ and $M_1=0$ & 100\% & $\chi^2_{\sum_m p_m}$                         &  52 & 145 & 0.95 & 0.95 & 0.97 & 0.97 \\ 
  &                                        &            & Hotelling's $T^2_{\sum_m p_m, N}$       &  64 & 156 & 0.95 & 0.95 & 0.97 & 0.98\\ 
  &                                        &            & Hotelling's $T^2_{\sum_m p_m, N-q-1}$ &  64 & 156 & 0.95 & 0.95 & 0.98 & 0.97 \\ 
  &                                        & 70\%   & $\chi^2_{\sum_m p_m}$                          &  74 & 206 & 0.95 & 0.95 & 0.97 & 0.98 \\ 
  &                                        &            & Hotelling's $T^2_{\sum_m p_m, N}$        &  86 & 218 & 0.95 & 0.95 & 0.98 & 0.97 \\ 
  &                                        &            & Hotelling's $T^2_{\sum_m p_m, N-q-1}$  &  86 & 218 & 0.95 & 0.95 & 0.98 & 0.97 \\
     \hline
\end{tabular}
\end{table}

\newpage

\subsection{Mis-specified error term distribution} 
 In both Tables \ref{Table: CP_MisSigma} and \ref{Table: CP_MisAR}, we observe that both the formulated and MC $CP$ estimates are close to $95\%$.
 Therefore, the proposed sample size calculator is robust against both the non-constant variance and the non-zero autocorrelation for the outcome measures over time.

\begin{table}[h]
\centering
\caption{
The $N$ and $CP$ estimates
when $\sigma_d^2$ is not constant over $d$.
}
\label{Table: CP_MisSigma}
\begin{tabular}{rlllrrrrrr}
  \hline
 & \multicolumn{3}{c}{} &  \multicolumn{6}{c}{Precision of average standardized proximal effect size}  \\
 & Trend of $\sigma_d^2$ & Availability & Test Statistics & 0.1 & 0.06 & 0.1 & 0.06 & 0.1 & 0.06 \\ 
 & \multicolumn{3}{c}{} &  \multicolumn{2}{c}{$N$} & \multicolumn{2}{c}{Formulated $CP$}  & \multicolumn{2}{c}{Monte Carlo $CP$} \\
  \hline
 & Increasing & 100\% & $\chi^2_{\sum_m p_m}$                          & 47 & 132 & 0.95 & 0.95 & 0.96 & 0.95 \\ 
 &                  &             & Hotelling's $T^2_{\sum_m p_m, N}$       &  59 & 143 & 0.95 & 0.95 & 0.95 & 0.95 \\ 
 &                  &             & Hotelling's $T^2_{\sum_m p_m, N-q-1}$ &  59 & 143 & 0.95 & 0.95 & 0.96 & 0.95 \\ 
 &                  & 70\%    & $\chi^2_{\sum_m p_m}$                          &  67 & 188 & 0.95 & 0.95 & 0.97 & 0.95 \\ 
 &                  &             & Hotelling's $T^2_{\sum_m p_m, N}$       &  79 & 199 & 0.95 & 0.95 & 0.96 & 0.95 \\ 
 &                  &             & Hotelling's $T^2_{\sum_m p_m, N-q-1}$ &  79 & 200 & 0.95 & 0.95 & 0.94 & 0.95 \\
  \\
& Decreasing & 100\% & $\chi^2_{\sum_m p_m}$                         &  47 & 132 & 0.95 & 0.95 & 0.94 & 0.94 \\ 
&                    &            & Hotelling's $T^2_{\sum_m p_m, N}$       &  59 & 143 & 0.95 & 0.95 & 0.95 & 0.96 \\ 
&                    &            & Hotelling's $T^2_{\sum_m p_m, N-q-1}$ &  59 & 143 & 0.95 & 0.95 & 0.96 & 0.96 \\ 
&                    & 70\%   & $\chi^2_{\sum_m p_m}$                          &  67 & 188 & 0.95 & 0.95 & 0.95 & 0.94 \\ 
&                    &            & Hotelling's $T^2_{\sum_m p_m, N}$       &  79 & 199 & 0.95 & 0.95 & 0.95 & 0.95 \\ 
&                    &            & Hotelling's $T^2_{\sum_m p_m, N-q-1}$ &  79 & 200 & 0.95 & 0.95 & 0.96 & 0.95 \\   
       \hline
\end{tabular}
\end{table}

\newpage

\begin{table}[h]
\centering
\caption{
The $N$ and $CP$ estimates
when $\epsilon_{id}=\phi\epsilon_{i, d-1} + \nu_{id}$, where $\phi$=$0.5$ and $-0.5$ and $\nu_{d}$ are i.i.d Normal(0, 0.75).
}
\label{Table: CP_MisAR}
\begin{tabular}{rlllrrrrrr}
  \hline
 & \multicolumn{3}{c}{} &  \multicolumn{6}{c}{Precision of average standardized proximal effect size} \\
 & $\phi$ & Availability & Test Statistics & 0.1 & 0.06 & 0.1 & 0.06 & 0.1 & 0.06 \\ 
 & \multicolumn{3}{c}{} &  \multicolumn{2}{c}{$N$} & \multicolumn{2}{c}{Formulated $CP$} & \multicolumn{2}{c}{Monte Carlo $CP$} \\
  \hline
 & $0.5$ & 100\% & $\chi^2_{\sum_m p_m}$                          & 47 & 132 & 0.95 & 0.95 & 0.96 & 0.95 \\ 
 &           &            & Hotelling's $T^2_{\sum_m p_m, N}$        & 59 & 143 & 0.95 & 0.95 & 0.95 & 0.94 \\ 
 &           &            & Hotelling's $T^2_{\sum_m p_m, N-q-1}$ & 59 & 143 & 0.95 & 0.95 & 0.95 & 0.96 \\ 
 &           & 70\%   & $\chi^2_{\sum_m p_m}$                          & 67 & 188 & 0.95 & 0.95 & 0.96 & 0.95 \\ 
 &           &            & Hotelling's $T^2_{\sum_m p_m, N}$        & 79 & 199 & 0.95 & 0.95 & 0.96 & 0.95 \\ 
 &           &            & Hotelling's $T^2_{\sum_m p_m, N-q-1}$  & 79 & 200 & 0.95 & 0.95 & 0.95 & 0.95 \\ 
  \\
  & $-0.5$ & 100\% & $\chi^2_{\sum_m p_m}$                         &  47 & 132 & 0.95 & 0.95 & 0.94 & 0.95 \\ 
  &            &            & Hotelling's $T^2_{\sum_m p_m, N}$       &  59 & 143 & 0.95 & 0.95 & 0.95 & 0.95 \\ 
  &            &            & Hotelling's $T^2_{\sum_m p_m, N-q-1}$ &  59 & 143 & 0.95 & 0.95 & 0.96 & 0.96 \\ 
  &            & 70\%   & $\chi^2_{\sum_m p_m}$                          &  67 & 188 & 0.95 & 0.95 & 0.95 & 0.94 \\ 
  &            &            & Hotelling's $T^2_{\sum_m p_m, N}$        &  79 & 199 & 0.95 & 0.95 & 0.95 & 0.95 \\ 
  &            &            & Hotelling's $T^2_{\sum_m p_m, N-q-1}$  &  79 & 200 & 0.95 & 0.95 & 0.95 & 0.95 \\ 
       \hline
\end{tabular}
\end{table}

\newpage

\subsection{Mis-specified trend of availability}\label{s:mistau precision}
In Table \ref{Table: CP_MisTau}, both the formulated and MC $CP$ estimates are close to the nominal $95\%$, showing that mis-specifying the availability rate trend over time has little impact on the $CP$, and consequently $N$ required. 

\begin{table}[h]
\centering
\caption{
The $N$ and $CP$ estimates
when $\tau_d$ is not constant over $d$, i.e. linearly increasing or decreasing over $d$.
}
\label{Table: CP_MisTau}
\begin{tabular}{rllrrrrrr}
  \hline
 & \multicolumn{2}{c}{} &  \multicolumn{6}{c}{Precision of average standardized proximal effect size}  \\
 & Trend of $\tau_d$ & Test Statistics & 0.1 & 0.06 & 0.1 & 0.06 & 0.1 & 0.06 \\ 
  & \multicolumn{2}{c}{} &  \multicolumn{2}{c}{$N$} & \multicolumn{2}{c}{Formulated $CP$}  & \multicolumn{2}{c}{Monte Carlo $CP$} \\
  \hline
& Increasing & $\chi^2_{\sum_m p_m}$                           &  67 & 188 & 0.95 & 0.95 & 0.94 & 0.95 \\ 
&                   & Hotelling's $T^2_{\sum_m p_m, N}$        &  79 & 199 & 0.95 & 0.95 & 0.95 & 0.96 \\ 
&                   & Hotelling's $T^2_{\sum_m p_m, N-q-1}$  &  79 & 200 & 0.95 & 0.95 & 0.96 & 0.96 \\
  \\
 & Decreasing & $\chi^2_{\sum_m p_m}$                          &  67 & 188 & 0.95 & 0.95 & 0.94 & 0.95 \\ 
 &                    & Hotelling's $T^2_{\sum_m p_m, N}$        &  79 & 199 & 0.95 & 0.95 & 0.95 & 0.94 \\ 
 &                    & Hotelling's $T^2_{\sum_m p_m, N-q-1}$  &  79 & 200 & 0.95 & 0.95 & 0.94 & 0.95 \\  
  \hline
\end{tabular}
\end{table}

\section{\texttt{R} implementation}\label{s: rimp}

\subsection{Implementation in \texttt{R} function}\label{s:demo}
\label{s:demo}
In this section, we demonstrate the implementation of the \texttt{R} function \texttt{SampleSize\_FlexiMRT} for sample size calculations via a simulation study. This function is available at \url{https://github.com/Kenny-Jing-Xu/FlexiMRT-SS/blob/master/SampleSizeFlexiMRT.R}. 

The inputs of the function are defined in the following.
\begin{verbatim}
SampleSize_FlexiMRT(days, occ_per_day, aa.day.aa, prob, 
        beta_shape, beta_mean, beta_initial, beta_quadratic_max, 
        tau_shape, tau_mean, tau_initial, tau_quadratic_max, 
        sigma, pow, sigLev, method, test, result, SS) 
\end{verbatim}
\begin{itemize}
\item \texttt{days}: the number of days during the study period
\item \texttt{occ\_per\_day}: the number of decision time points per day during the study period. Note it is fixed at \texttt{occ\_per\_day=1} for DIAMANTE study
\item \texttt{aa.day.aa}: the day of each proposed intervention category
\item \texttt{prob}: the allocation probability matrix of the levels, with a dimension of D by (M+1), where the first column is the allocation probabilities of control level
\item \texttt{beta\_shape}: the shape of $b_m(d; \boldsymbol\beta_m)$ in terms of $d$, i.e. ``constant'', ``linear'', ``linear and constant'', or ``quadratic"
\item \texttt{beta\_mean}: the average of the standardized proximal effect size of each intervention category over the study period
\item \texttt{beta\_initial}: the initial standardized proximal effect size of each intervention category
\item \texttt{beta\_quadratic\_max}: the first value of $d$ that gives the turning value of $b_m(d; \boldsymbol\beta_m)$ if the shape is ``quadratic'' or ``linear and constant''.
\item \texttt{tau\_shape}: the shape of $\tau_{d}$ in terms of $d$, i.e. ``constant", ``linear", ``linear and constant", or ``quadratic"
\item \texttt{tau\_mean}: the average of $\tau_{d}$ over the study period
\item \texttt{tau\_initial}: the initial $\tau_{d}$ 
\item \texttt{tau\_quadratic\_max}: the value of $d$ that gives the maximum value of $\tau_{d}$ if the shape is quadratic
\item \texttt{sigma}: $\sigma$ or the standard deviation of the error term $\epsilon$
\item \texttt{pow}: the power
\item \texttt{sigLev}: the type-I error rate
\item \texttt{method}: the method of sample size calculation, based on either power, i.e. ``power" or confidence interval, i.e. ``confidence interval"
\item \texttt{test}: the test statistics, based on Chi-squared distribution, i.e. ``chi", Hotelling's T-squared distribution with $N-q-1$ degree of freedom, i.e. ``hotelling N-q-1",  Hotelling's T-squared distribution with $N-1$ degree of freedom, i.e. ``hotelling N-1", or Hotelling's T-squared distribution with $N$ degree of freedom, i.e. ``hotelling N"    
\item \texttt{result}: the chosen calculated result, i.e ``choice\_sample\_size", ``choice\_power" or ``choice\_coverage\_probability"
\item \texttt{SS}: the specified sample size if the result of either power or coverage probability is chosen 
\end{itemize}

The possible outputs are summarized below:
\begin{itemize}
\item \texttt{N}, the estimated sample size;
\item \texttt{P}, the estimated power;
\item \texttt{CP}, the estimated coverage probability;
\item \texttt{d}, the consistently estimated $\boldsymbol\beta$;
\item \texttt{Sig\_bet\_inv}, the inverse of the asymptotic covariance estimated for $\boldsymbol\beta$.
\end{itemize}

Here we present the \ts{R} code for the sample size  calculation corresponding to 
\begin{itemize}
\item Test statistics=Hotelling's $T^2_{Mp, N-q-1}$,
\item $M=4$, where $M=M_0+M_1$, $M_0=3$ and $M_1=1$,
\item $D=180$,
\item average proximal standardized effect=$0.1$ for each intervention category,
\item proximal effects increased linearly at the beginning and maintained constant after 28-day,
\item constant availability rate=$70\%$.
\end{itemize} 


\begin{verbatim}
# One decision time point per day during the study period 
occ_per_day=1 

# The number of parameters of beta for each intervention category, i.e. p_m
pb=2

# The study period of 180-day
days=180

# Three intervention categories are proposed at the beginning and 
# another one is proposed in the middle of the trial.
aa.each=c(3, 1)

# The first day, the middle day and the last day of the study period
aa.day=c(1, ( floor(days/2) + 1 ), days)

# The number of days when the intervention categories are proposed in the trial
aa.freq=length(aa.day)-1

# The days that each of the intervention categories are added to the trial, 
# e.g., c(1,1,1,91)  
aa.day.aa=rep( ( aa.day )[1:aa.freq], aa.each )

# The allocation probability matrix 
# Based on uniform random with dimension 180 by 5 
# The first column is the allocation probability of the control 
# category. 
# Other columns are the probabilities of intervention 
# categories.
# Rows 1-90 are c(0.25, 0.25, 0.25, 0.25, 0).
# Rows 91-180 are c(0.2, 0.2, 0.2 ,0.2, 0.2).
prob = matrix( 0, days, ( 1 + sum(aa.each) ) )
for( l in 1:( length( aa.day ) - 1 ) )
{
  prob[ ( aa.day[ l + 0 ] + 0 ):( aa.day[ l + 1 ] - 1 ), 
        1:( 1 + cumsum( aa.each )[ l ] ) ] <- 
              1/( cumsum( aa.each )[ l ] + 1 )
}
prob[ ( aa.day[ length( aa.day ) ] ), 
      1:( 1 + sum( aa.each ) ) ] <- 
            1/( 1 + sum( aa.each ) )
\end{verbatim}


\begin{verbatim}
# Linear until 28-day then constant in terms of d for the beta shape
beta_shape= "linear and constant"
# Average proximal standardised effect size for each intervention category
beta_mean=rep(0.1, sum( aa.each ) ) 
# Initial proximal standardised effect size for each intervention category
beta_initial=rep(0.01, sum( aa.each ) )
# Maximum proximal standardised effect day for each intervention category
beta_quadratic_max=aa.day.aa-1+28

# Constant shape for availability proportion
tau_shape="constant" 
# Average availability proportion
tau_mean=0.7
# Initial availability proportion
tau_initial=0.7
# Maximum availability day
tau_quadratic_max=28

# The power 0.8 and type-I error rate 0.05
pow=0.8 
sigLev=0.05

# The standard deviation and correlation coefficient of the error term
sigma=1
rho=0

# The sample size calculated  by power and test statistics 
# with the distribution of Hotelling's T-squared with degree of freedom N-q-1
method="power"
test = "hotelling N-q-1"
result = "choice_sample_size"

MRTN=SampleSize_FlexiMRT(days=days, occ_per_day=occ_per_day, 
                     aa.day.aa = aa.day.aa, 
                     prob=prob, 
                     beta_shape=beta_shape, 
                     beta_mean=beta_mean, 
                     beta_initial=beta_initial, 
                     beta_quadratic_max=beta_quadratic_max, 
                     tau_shape=tau_shape, tau_mean=tau_mean, 
                     tau_initial=tau_initial, 
                     tau_quadratic_max=tau_quadratic_max, 
                     sigma=sigma, pow=pow, sigLev=sigLev, 
                     method=method, test=test, 
                     result=result)
                     
# The calculated sample size
N=MRTN$N
\end{verbatim}
                     


The above R codes calculate sample size $73$.  The formulated power achieved under this number of participants ($0.80$) can be calculated by substituting the inputs of
\texttt{result="choice\_power"}
and
 \texttt{SS=73} 
in the \texttt{R} function \texttt{SampleSize\_FlexiMRT}.

\subsection{Implementation using \texttt{R} shiny}\label{s: Rshiny}
\begin{figure}
\begin{center}
\begin{tabular}{cc}
\includegraphics[height=4cm, width=8cm]{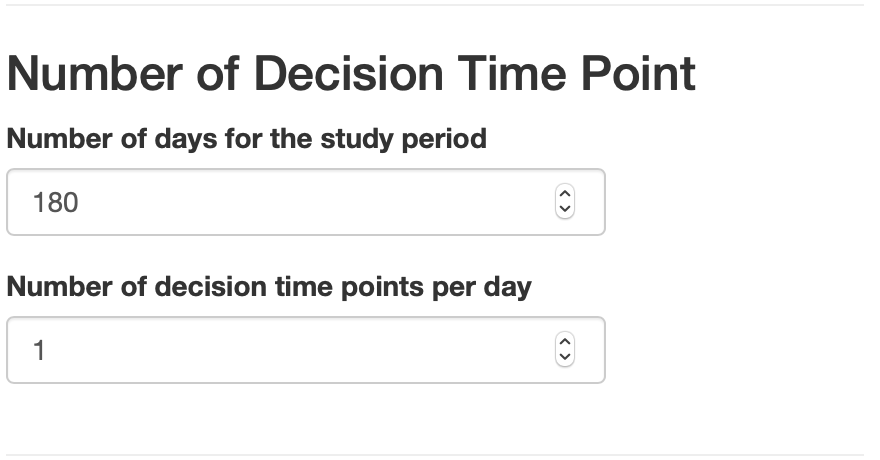} & \includegraphics[height=4cm, width=8cm]{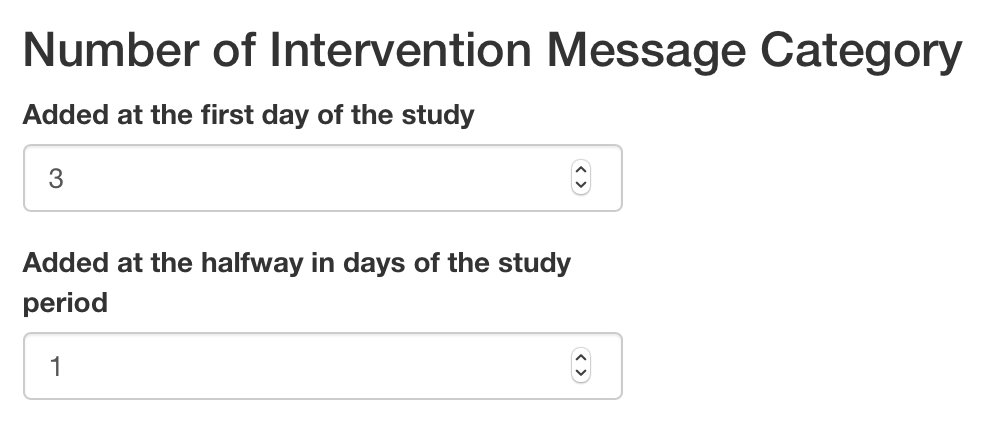} \\
 a) Number of decision time point & b) Number of intervention message category \\
  \includegraphics[height=4cm, width=8cm]{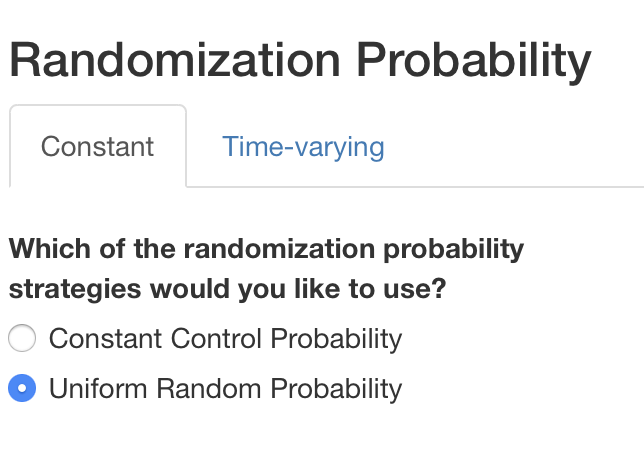} & \includegraphics[height=4cm, width=8cm]{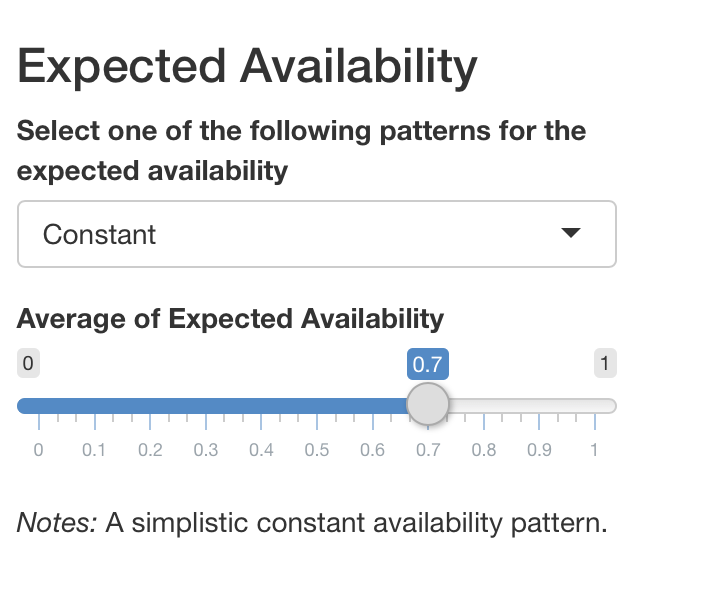} \\
 c) Randomisation probability & d) Expected availability \\
  \includegraphics[height=3cm, width=8cm]{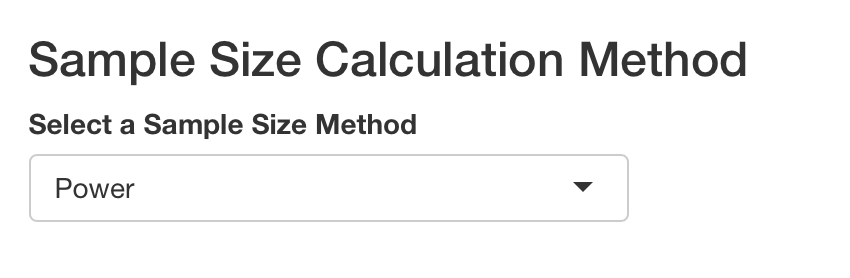} & \includegraphics[height=3cm, width=8cm]{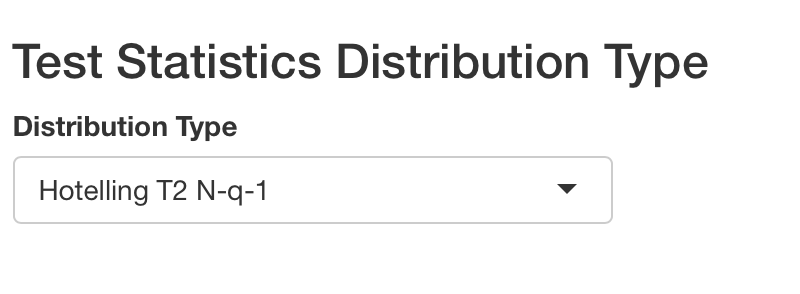} \\
 e) Sample size calculation method & f) Test statistics distribution type \\
  \includegraphics[height=5cm, width=6cm]{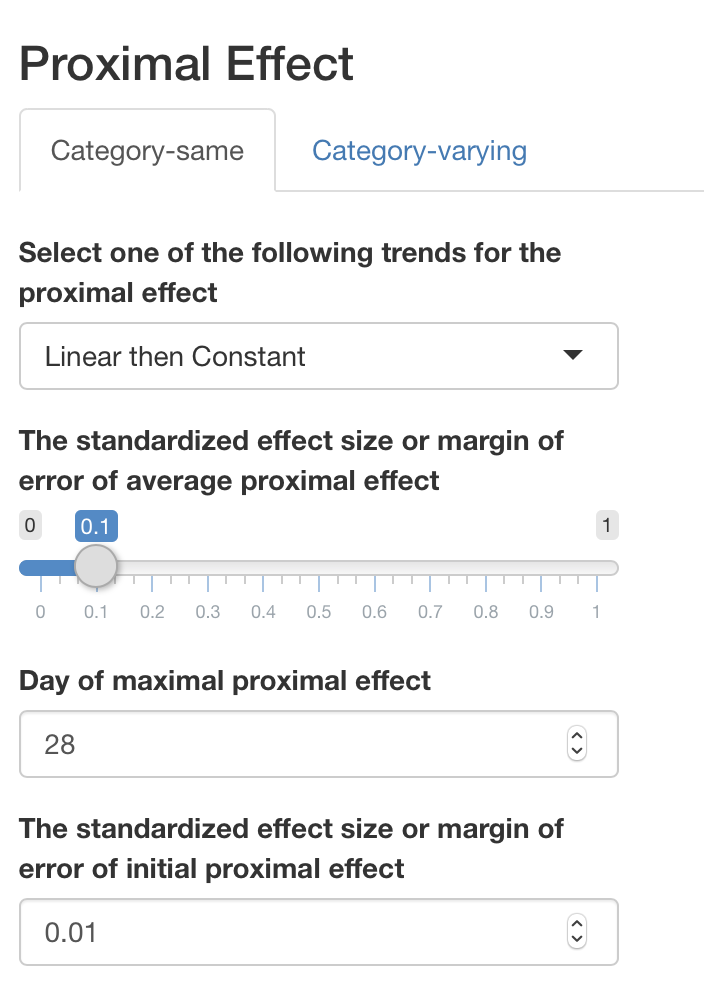} &  \includegraphics[height=5cm, width=6cm]{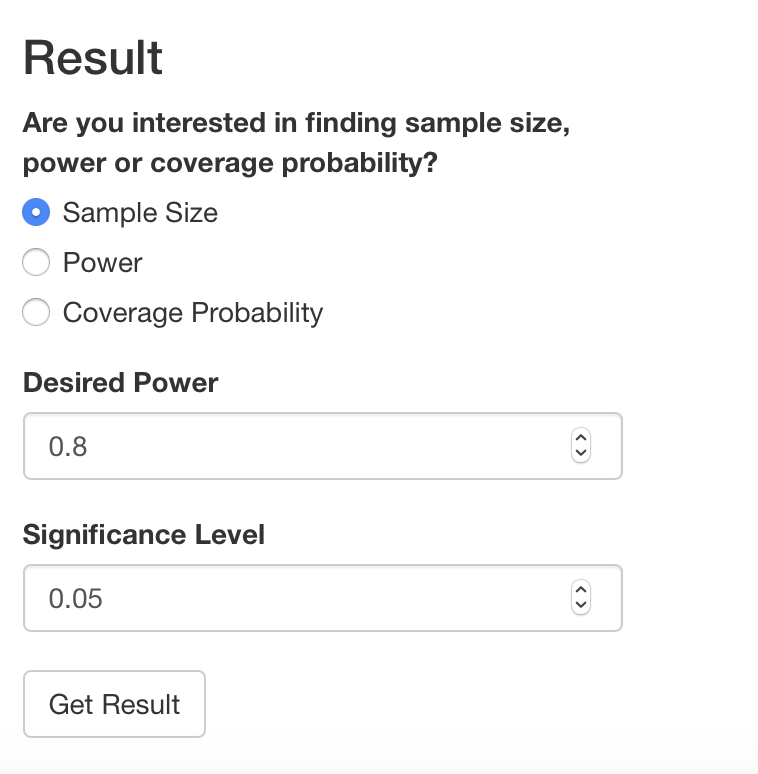} \\
 g) Proximal effect & h) Result
\end{tabular}
\end{center}
\caption{
a) The study duration is $180$ days with $1$ decision time point a day. b) The number of intervention category added at the first day is $3$ and added half-way through the study in days is $1$. c) The randomization probability is based on uniform random intervention. d) The expected availability of each participant at each decision time point is $70\%$. e) The sample size calculation method is based on power. f) The test statistics is Hotelling's $T^2$ distributed with denominator degrees of freedom $N-q-1$. g) The proximal effect is increased linearly until 28-day then maintaining constantly at its maximum value. The initial and average values of the standardised proximal effect size for any of the intervention levels are $0.01$ and $0.1$ respectively.  h) The required sample size is calculated at a nominal power of $80\%$ and level of significance $5\%$. 
}\label{fig: shiny}
\end{figure}

Similar to the online sample size calculator (MRT-SS) for the micro-randomized trial created by Nicholas \textit{et al.} \cite{Nicholas_etal_2016}, we create a proposed sample size calculator named ``FlexiMRT-SS'' using R shiny.
The web link to this application is \url{https://kennyxu.shinyapps.io/FlexiMRT-SS/}. 
We explore each of its components briefly using Figure \ref{fig: shiny} based on the same example in Section \ref{s:demo}. 

Figure \ref{fig: shiny} describes a trial that has a study period of $180$ days with only one decision time point per day. For a particular component, three intervention categories are proposed before the first day of study, and another one is added halfway through the study, i.e. after $90$ days.
 At each decision time point, a participant is randomly allocated to either the control category or one of the active intervention categories based on an uniform random intervention strategy. 
 For example, the randomization probability for each category (control or intervention) is 1/4 for the first half of the study period, while the randomization probability for each level is $0.2$ for the second half of the study period when one more category is added.  
The availability of each participant is expected $70\%$ at all the decision time points. 
The sample size calculation method is based on power. The test statistic is assumed to follow a Hotelling's $T^2$ distribution with denominator degrees of freedom $N-q-1$.
For the proximal effect, it is assumed to follow a trend of increasing linearly from day-1 to day-28 and then maintaining constantly at its maximum value. The initial and average values of the standardized proximal effect size for either intervention level are $0.01$ and $0.1$ respectively.
In the `Result' section, the sample size is calculated as the final output under a nominal $80\%$ power and a level of significance of $0.05$. The output is presented by ``The required sample size is 73 to attain 80\% power when the significance level is 0.05.'' after the `Get Result' button is clicked. 
Alternatively, if `Power' is selected in the `Result' section, and $73$ is specified in the `Number of Participants' cell, the output is going to be ``The sample size 73 gives 80\% power when the significance level is 0.05''.

\end{document}